%% file: double_arxiv.tex
\documentclass[final,twcolumn,10pt]{IEEEtran}
\usepackage[dvips,final]{graphicx}
\usepackage{amssymb}    
\usepackage{cite}
\usepackage[cmex10]{amsmath}
\usepackage{verbatim}

\input ajmacros

\begin{document}
\newtheorem{assu}{Assumption}
\newtheorem{prop}{Property}
\newtheorem{rema}{Remark}

\title{UWB Signal Detection by Cyclic Features}

\author{\normalsize \IEEEauthorblockN{Yiyin Wang\IEEEauthorrefmark{2}\IEEEauthorrefmark{1}, Xiaoli Ma\IEEEauthorrefmark{3}, Cailian Chen\IEEEauthorrefmark{2}, and Xinping Guan\IEEEauthorrefmark{2}}\\
\IEEEauthorblockA{\IEEEauthorrefmark{2}Department of Automation, Shanghai Jiao Tong University, Shanghai, 200240, P. R. China}\\
\IEEEauthorblockA{\IEEEauthorrefmark{3}\normalsize School of ECE, Georgia Institute of Technology,
Atlanta, GA 30332-0250, USA}
\thanks{Part of the work was supported by the National Nature Science Foundation of China (No. 61301223), the Nature Science Foundation  of Shanghai (No. 13ZR1421800), and the Georgia Tech Ultra-wideband Center of Excellence (http://www.uwbtech.gatech.edu/). Some preliminary results of this work were presented at the IEEE International Conference on Ultra-Wideband, Sydney, Australia, September 2013}}

\maketitle

\begin{abstract}
Ultra-wideband (UWB) impulse radio (IR) systems are well known for low transmission power, low probability of detection, and overlaying with narrowband (NB) systems. These merits in fact make UWB signal detection challenging, since several high-power wireless communication systems coexist with UWB signals. In the literature, cyclic features are exploited for signal detection. However, the high computational complexity of conventional cyclic feature based detectors burdens the receivers. In this paper, we propose computationally efficient detectors using the specific cyclic features of UWB signals. The closed-form relationships between the cyclic features and the system parameters are revealed. Then, some constant false alarm rate detectors are proposed based on the estimated cyclic autocorrelation functions (CAFs). The proposed detectors have low complexities compared to the existing ones. Extensive simulation results indicate that the proposed detectors achieve a good balance between the detection performance and the computational complexity in various scenarios, such as multipath environments, colored noise, and NB interferences.
\end{abstract}


EDICS: SSP-CYCS Cyclostationary signal analysis, SSP-DETC Detection, SPC-UWBC Ultra wideband communications

\section{Introduction}

Ultra wideband technology is adopted by the IEEE 802.15.4a standard \cite{S802.15.4a} for wireless personal area networks (WPANs) to provide low-power communications and precise ranging capabilities. It is featured by sharing the spectrum with other communication systems to efficiently use rare spectrum resources \cite{Hamalainen2002,Giuliano2005,Chiani2009,Pinto2009}. For example, according to the IEEE 802.15.4a standard, the specified UWB signal may occupy the same spectrum as the signal specified by the IEEE 802.16 standard \cite{S802.16} (also named as worldwide interoperability for microwave access (WiMAX)). Therefore, UWB systems usually work in a heterogeneous wireless environment. The first fundamental task of a UWB receiver is to detect the transmitted UWB signal regardless of interferences in heterogeneous environments. Conventional energy detectors fail to distinguish different signals from each other. Moreover, UWB channel environments are rich in multipaths and subject to varying noise. Hence, the detectors based on known signal and noise statistics, such as matched filters, are impractical for implementation.

Recently, cyclostationarity is of interest for detecting the signal in cognitive radio (CR) systems \cite{Oner2007,Dobre2007,Yucek2009,Ma2009,Zeng2010}, in which secondary users detect the presence of primary users and make use of the unoccupied spectrum efficiently. Cyclostationarity describes the periodicity of the statistical properties of a signal, and exists in almost all modulated signals \cite{Gardner1991,Giannakis1998,Antoni2007,Napolitano2013}. The signal parameters, e.g. modulation types, symbol rates, carrier frequencies, and periods of spreading codes, determine cyclic features of a signal. Since these parameters are different for different types of signals, the distinct cyclic features can be exploited for signal detection. In addition, the cyclic feature detectors are robust to noise uncertainty. For CRs, the Dandawate-Giannakis detector \cite{Dandawate1994} is employed by secondary users to detect various primary signals, such as orthogonal frequency division multiplexing (OFDM) signals, Gaussian minimum shift keying (GMSK) modulated signals, and code division multiple access (CDMA) signals \cite{Oner2007,Oner2007a,Punchihewa2010}. A multi-cycle extension of the Dandawate-Giannakis detector as well as its computationally efficient modifications are proposed to detect the OFDM signals in \cite{Lunden2009}. Consequentially, a collaborative detection among secondary users with censoring is also developed in \cite{Lunden2009}. Furthermore, a multi-antenna extension of the Dandawate-Giannakis detector is designed in \cite{Zhong2010} to take advantage of the diversity gain of multiple antennas.

Since the computational complexities of the Dandawate-Giannakis detector and its inheritors are relatively high, several heuristic cyclic detectors considering noise uncertainty are designed in \cite{Tani2010,Bogale2012,Huang2013} to reduce the complexity. A single-cycle single-lag detector is proposed in \cite{Tani2010} to detect OFDM signals of WiMAX systems. A multi-cycle single-lag detector is further developed in \cite{Bogale2012} to perceive the OFDM signals. Taking colored Gaussian noise into account, a multi-cycle multi-lag cyclic feature detector is derived in \cite{Huang2013}, and it can accommodate multiple antennas as well. Although some of these proposed detectors can be adapted for UWB signals, such as the ones in \cite{Lunden2009,Huang2013,Wang2013a}, they either maintain high complexity or do not take advantage of the unique properties of UWB signals. For example, a Dandawate-Giannakis type detector is employed in \cite{Wang2013a} to detect a UWB signal under the coexistence of a GMSK signal. However, it still suffers from high computational complexity. Furthermore, a wide band spectrum sensing method based on recovered sparse 2-D cyclic spectrum from compressive samples is proposed in \cite{Tian2012}. This sub-Nyquist scheme is attractive, as the Nyquist rate of the UWB signal is notably high. However, a 2-D cyclic spectrum recovery is not necessary here, as some prior knowledge of the cyclic features of the UWB signal has been assumed. Moreover, a detection and avoidance (DAA) scheme is proposed in \cite{Mishra2007, Tani2010} to facilitate the coexistence of UWB systems and WiMAX systems, where the UWB devices as secondary users are able to detect the presence of WiMAX systems by their cyclic features, and avoid the transmission in the occupied spectrum. Different from \cite{Mishra2007, Tani2010}, our work focuses on the receiver to detect the UWB signal in heterogeneous environments, not on the transmitter to sense the available spectrum.

As a result, multi-cycle multi-lag detectors based on cyclic features are proposed for UWB receivers to recognize the UWB signals of interest in this paper. At first, the UWB signal structure is specified by following the IEEE 802.15.4a standard. Sequentially, the cyclic features of the UWB signal are investigated. Although the cyclic features of non-standard UWB signals have also been analyzed in \cite{Oner2008,Vucic2009,Wang2013a}, where the symbol rate plays the main role, due to a hybrid modulation and scrambling codes, the cyclic features of the standard UWB signal do not simply appear at consecutive multiples of the symbol rate. The closed-form relationships between the cyclic features and the system parameters are further revealed. Furthermore, constant false alarm rate (CFAR) detectors are proposed based on the estimated cyclic autocorrelation functions (CAFs). Thanks to the ultra wide bandwidth of the signal and the resolvable multipath channel components, the proposed detectors take advantage of multiple cyclic frequencies (CFs) and multiple time lags (TLs). Their computational complexities are significantly less than the ones of the Dandawate-Giannakis type detectors in \cite{Lunden2009}, and are comparable to the complexity of the detector proposed in \cite{Huang2013}, which deals with colored Gaussian noise. Note that the detector \cite{Huang2013} fails under the case of interference corruption, as it oversimplifies the covariance estimation. On the other hand, the proposed detector composed of the single-cycle single-lag Dandawate-Giannakis test statistics can still deal with the interferences, and it achieves a tradeoff between detection performance and computational complexity.



The rest of the paper is organized as follows. The preliminary knowledge of the cyclostationarity is reviewed in Section~\ref{sec:pre}. The model of the IEEE 802.15.4a UWB signal and its cyclic features are described and analyzed, respectively, in Section~\ref{sec:model}. Consequentially, CFAR detectors are developed based on the estimated CAFs to exploit the specific cyclic features of the UWB signal in Section~\ref{sec:CF_det}. The comparison of the computational complexities between the proposed detectors and the existing ones is carried out in Section~\ref{sec:complx}. Extensive simulations validate the detection performance under various scenarios in Section~\ref{sec:sim}. The conclusions are drawn in Section~\ref{sec:concl}.

\section{Preliminaries of cyclostationarity}\label{sec:pre}

In this section, we briefly review the preliminary knowledge for cyclostationary processes and introduce the notations. More comprehensive details can be found in \cite{Gardner1991,Giannakis1998,Antoni2007,Napolitano2013}.

A cyclostationary process is characterized by the cyclically varied statistical properties of a signal with respect to (w.r.t.) time. A special case of cyclostationary signals is the wide-sense cyclostationary signal, whose second-order statistics is periodic in time. Hence, the autocorrelation function of a zero-mean wide-sense cyclostationary signal $s(t)$ is given by
\beqa
\Upsilon_{ss}(t,\tau)\! = \!E[s(t+\tau/2)s^\ast(t-\tau/2)] \!=\!\! \Upsilon_{ss}(t + nT_f,\tau), \label{cov_nc}\\
\forall n \in \mathcal{Z}\nonumber
\enqa where $\ast$ denotes the complex conjugate, $\mathcal{Z}$ is the integer set, $\tau$ is the time lag (TL), and $T_f$ is the fundamental period. As a result, $\Upsilon_{ss}(t,\tau)$ can be decomposed into Fourier series as
\beqa
\Upsilon_{ss}(t,\tau) = \sum_{\alpha_k \in \mathcal{A}}R_{ss}(\alpha_k,\tau)e^{j2\pi \alpha_k t},
\enqa where $\mathcal{A} = \{\alpha_k| R_{ss}(\alpha_k,\tau)\neq 0 \}$ is the set of cyclic frequencies (CFs), and $\alpha_k$ is related to the fundamental period as $\alpha_k = k/T_f, k \in \mathcal{Z}$. The Fourier coefficients $R_{ss}(\alpha_k,\tau)$ can be calculated as
\beqa
R_{ss}(\alpha_k,\tau) = \lim_{T \rightarrow \infty}\frac{1}{T}\int_{-T/2}^{T/2}\Upsilon_{ss}(t,\tau)e^{-j2\pi \alpha_k t}dt,
\enqa which is also named the cyclic autocorrelation function (CAF), and $T$ is the observation period. Consequently, the spectrum correlation density (SCD) function is defined as the Fourier transform of the CAF
\beqa
S_{ss}(\alpha_k,f) = \int_{-\infty}^{\infty} R_{ss}(\alpha_k,\tau)e^{-j2\pi f\tau} d\tau.
\enqa The counterparts of these functions are the conjugate ones. Let us start with the conjugate autocorrelation function given by
\beqa
\Upsilon_{ss^\ast}(t,\tau)\! = \!E[s(t+\tau/2)s(t-\tau/2)]. \label{cov_c}
\enqa Its Fourier coefficients are the conjugate CAF $R_{ss^\ast}(\alpha_k,\tau)$. Hence, the Fourier transform of $R_{ss^\ast}(\alpha_k,\tau)$ is the conjugate SCD $S_{ss^\ast}(\alpha_k,f)$. To combine all these definitions, we denote them as $\Upsilon_{ss^{\left({}^{}_{\ast}\right)}}(t,\tau), R_{ss^{\left({}^{}_{\ast}\right)}}(\alpha_k,\tau)$ and $S_{ss^{\left({}^{}_{\ast}\right)}}(\alpha_k,f)$, where $\left({}^{}_{\ast}\right)$ represents two options (nonconjugate and conjugate).



\section{The IEEE {802.15.4a} UWB Signal Model and Its Cyclic Features}\label{sec:model}

\subsection{Signal Model}

\begin{table*}[t]
\centering
\begin{tabular}{c l}
$p(t)$ & the transmitted UWB pulse of length $T_p$\\
$\epsilon$ & the unknown deterministic timing offset    \\
$N_{cpb}$ & the number of chips per burst \\
$N_{burst}$ & the number of burst per symbol\\
$N_{hop}$ & the number of hopping burst per symbol\\
  $T_c$ & the chip interval \\
   $T_{burst}$ & the burst interval, where $T_{burst} = N_{cpb}T_c$, \\
   $T_{BPM}$ & the position shift for the BPM, where $T_{BPM} = T_{dsym}/2$  \\
   $T_{dsym}$ & the symbol period, where $T_{dsym} = N_{burst}T_{burst}$, \\
   $a_k$& the $k$th symbol modulates the burst amplitude, where $a_k \in \{ \pm 1 \}$\\
   $b_k$& the $k$th symbol modulates the burst position, where $b_k \in \{0, 1 \}$\\
   $c_{n+kN_{cpb}}$& the scrambling code for the $k$th symbol, where $c_{n+kN_{cpb}} \in \{\pm 1\}, n = 1,\dots,N_{cpb}$\\
   $h^{(k)}$& the burst hopping code for the $k$th symbol, where $h^{(k)} \in \{0,1,\dots, N_{hop}-1\}$
\end{tabular}
\caption{Parameters for the IEEE 802.15.4a UWB signal \cite{S802.15.4a} in (\ref{UWB_model})}\label{UWB_para}
\end{table*}

\begin{figure*}
\centering
\centerline{\includegraphics[width=0.85\linewidth]{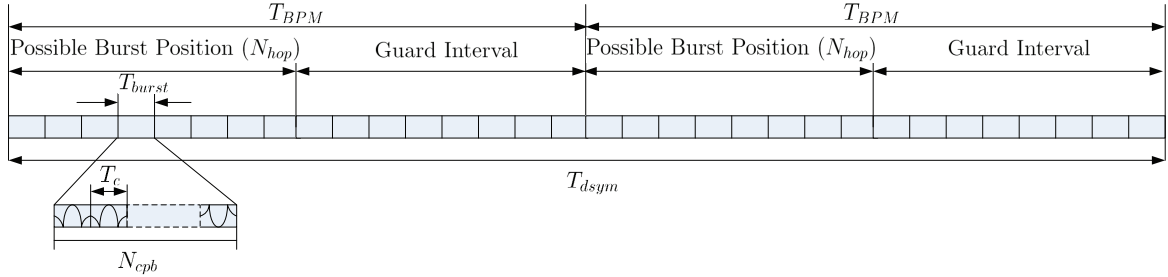}}
\caption{The UWB symbol structure according to the IEEE 802.15.4a standard \cite{S802.15.4a}}\label{UWB_symbol}
\end{figure*}

According to the IEEE 802.15.4a standard \cite{S802.15.4a}, the equivalent baseband model for the UWB PHY transmitting signal can be written as
\beqa\label{UWB_model}
x(t)\!\!&\!\!=\!\!&\!\! \!\!\sum_{k=-\infty}^{+\infty}a_k \!\!\sum_{n=0}^{N_{cpb}-1}c_{n+kN_{cpb}} \\
&\!\!\!\!&\times p(t-kT_{dsym}-b_kT_{BPM}-h^{(k)}T_{burst}-nT_c-\epsilon),\nonumber
\enqa which is modulated by a combination of burst position modulation (BPM) and binary phase-shift keying (BPSK). The parameters and notations in (\ref{UWB_model}) are listed in Table \ref{UWB_para}. The UWB symbol structure is shown in Fig. \ref{UWB_symbol}, where a UWB symbol is composed of a burst of UWB pulses, whose amplitudes are modulated by the data symbol $a_k$ and the scrambling sequence $c_{n+kN_{cpb}}$, and whose positions are modulated by the data symbol $b_k$ and the burst hopping sequence $h^{(k)}$. Note that $c_{n+kN_{cpb}}$ and $h^{(k)}$ are different for each symbol. They facilitate the multiuser interference suppression and spectral smoothing, and are generated from a common pseudo-random bit stream (PRBS) scrambler. Consequentially, the scrambling sequence is given by \cite{S802.15.4a}
\beqa
c_n = c_{n-14} \oplus c_{n-15},
\enqa where $\oplus$ denotes modulo-two addition, and the hopping sequence is generated as \cite{S802.15.4a}
\beqa
h^{(k)}=2^0c_{n+kN_{cpb}} +\dots+2^{m-1}c_{m-1+kN_{cpb}}, 
\enqa where $m= {\rm log}_2(N_{hop})$. Since the polynomial order of the PRBS scrambler is high and $m$ takes values from $\{1,3,5\}$, the scrambling sequence $c_{n+kN_{cpb}}$ and the hopping sequence $h^{(k)}$ can be assumed as wide-sense stationary (WSS) and independent from each other. Furthermore, $a_k$ and $b_k$ are also assumed to be WSS and independent from each other. Therefore, $c_{n+kN_{cpb}}$ and $a_k$ take values from $\{\pm 1\}$ with equal probability, and $b_k$ selects values from $\{0,1\}$ with equal probability as well. Moreover, the hopping sequence $h^{(k)}$ chooses values from $\{0,1,\dots, N_{hop}-1\}$ with equal probability.

\subsection{Cyclic Features of IEEE {802.15.4a} UWB Signals}\label{sec:CF}
In this subsection, we investigate the cyclic features of the IEEE 802.15.4a UWB signal $x(t)$ in (\ref{UWB_model}). As BPSK and scrambling code are employed, $x(t)$ is a zero-mean random signal. Since $x(t)$ is a real signal, $\Upsilon_{xx}(t,\tau)$ and $\Upsilon_{xx^\ast}(t,\tau)$ are equivalent in this case. Without loss of generality, $\Upsilon_{xx}(t,\tau)$ is derived by plugging (\ref{UWB_model}) into (\ref{cov_nc}) as (see Appendix \ref{app:AF} for more details)
\beqa\label{ACF_UWB}
{\Upsilon}_{xx}(t,\tau)
&\!\!\!\!=\!\!\!\!&\alpha_1^x\sum_{q=-\infty}^{+\infty} e^{j2\pi \alpha_q^x(t-\epsilon)}\bar{\beta}(\alpha_q^x)\bar{\eta}(\alpha_q^x)\nonumber\\
&& \times \phi_p(\alpha_q^x,\tau) w\left(T_c\alpha_q^x,N_{cpb}\right),
\enqa where $\alpha_q^x = q/T_{dsym}, q\in \mathcal{Z}$, $\bar{\beta}(f) = E[e^{-j2\pi  b_k T_{BPM}f}]$, $\bar{\eta}(f) = E[e^{-j2\pi h^{(k)} T_{burst}f}]$,
\beqa\label{eq:phi_p1}
\phi_p(\alpha_q^x,\tau) &=& \int P(z+\alpha_q^x)P^\ast(z)  e^{j2\pi \tau (z+\alpha_q^x/2)}  dz\nonumber\\
 &=& p(\tau)e^{-j2\pi\tau\alpha_q^x/2} \otimes p^\ast(-\tau)e^{j2\pi\tau\alpha_q^x/2},
\enqa with $\otimes$ denoting convolution, $P(f)$ being the Fourier transform of $p(t)$, and
\beqa
w(\rho,H)= \sum_{n=0}^{H-1} e^{-j2\pi \rho n }=\displaystyle\frac{{\rm sin}(\pi \rho H)}{{\rm sin}(\pi \rho)}e^{-j\pi \rho (H-1)}.\enqa
In (\ref{ACF_UWB}), $\Upsilon_{xx}(t,\tau)$ is decomposed into the Fourier series using $1/T_{dsym}$ ($\alpha_1^x$) as the fundamental CF w.r.t. $t$. The Fourier coefficient of $\Upsilon_{xx}(t,\tau)$ is the CAF $R_{xx}(\alpha_q^x,\tau)$. Recall that $b_k$ and $h^{(k)}$ select values from $\{0,1\}$ and $\{0,1,\dots, N_{hop}-1\}$ with equal probability, respectively, and they are independent with each other. The CAF $R_{xx}(\alpha_q^x,\tau)$ is simplified as
\beqa\label{CAF_UWB}
R_{xx}(2\alpha_q^x,\tau)
&\!=\!&\frac{\displaystyle \alpha_1^x }{\displaystyle N_{hop}}w\left(\frac{2q}{N_{burst}N_{cpb}},N_{hop}N_{cpb}\right)\nonumber\\
&&\times \phi_p(2\alpha_q^x,\tau)e^{-j4\pi\alpha_q^x\epsilon}, q \in \mathcal{Z}.
\enqa Please refer Appendix \ref{app:CAF} for more details about this derivation. Several remarks are due here.
\begin{rema}
The function $\displaystyle w\left(2q/(N_{burst}N_{cpb}),N_{hop}N_{cpb}\right)$ reaches local maxima, when $q = k N_{burst}N_{cpb}/2$ with $k \in \mathcal{Z}$. Meanwhile it equals zero, when $\displaystyle q = k^\prime N_{burst}/(2N_{hop})$ with $k^\prime \in \mathcal{Z}$ and $k^\prime \neq kN_{hop}N_{cpb}/2$. Therefore, there are $N_{cpb}N_{hop} - 1$ zeros between two adjacent local maximum values of $\displaystyle w\left(2q/(N_{burst}N_{cpb}),N_{hop}N_{cpb}\right)$. As a result, the nonzero pattern of $\displaystyle w\left(2q/(N_{burst}N_{cpb}),N_{hop}N_{cpb}\right)$ w.r.t. $q$ is related to the product $N_{burst}N_{cpb}/2$ and $N_{burst}/(2N_{hop})$.
\end{rema}
\begin{rema}
According to (\ref{CAF_UWB}), the nonzero pattern of $R_{xx}(2\alpha_q^x,\tau)$ w.r.t. $q$ is determined by $\displaystyle w\left(2q/(N_{burst}N_{cpb}),N_{hop}N_{cpb}\right)$, and thus also related to the product $N_{burst}N_{cpb}/2$ and $N_{burst}/(2N_{hop})$. Moreover, the ranges of the nonzero support of $R_{xx}(2\alpha_q^x,\tau)$ w.r.t. $q$ and $\tau$ are all determined by $\phi_p(2\alpha_q^x,\tau)$, which is related to the bandwidth of the UWB pulse.
\end{rema}

\begin{rema}
Although we do not take the multipath channel into the derivation, its impact on the CAF can be analyzed. Let us assume the multipath channel $h(t)$ does not change during the detection. Due to the multipath channel, the received pulse shape is that $\widetilde{p}(t) = p(t) \otimes h(t)$. Hence, the CAF of the UWB signal via a multipath channel $\widetilde{R}_{xx}(2\alpha_q^x,\tau)$ is still given by (\ref{CAF_UWB}), but replacing $\phi_p(\alpha_q^x,\tau)$ with $\widetilde{\phi}_p(\alpha_q^x,\tau)$, which is as follows
\beqa\label{eq:phi_p3}
\widetilde{\phi}_p(\alpha_q^x,\tau) = \int \widetilde{P}(z+\alpha_q^x)\widetilde{P}^\ast(z)  e^{j2\pi \tau (z+\alpha_q^x/2)} d z,
\enqa where $\widetilde{P}(f) = H(f)P(f)$ and $H(f)$ is the Fourier transform of the multipath channel $h(t)$. As a result, the nonzero pattern of $\widetilde{R}_{xx}(2\alpha_q^x,\tau)$ is still decided by $w\left(2q/(N_{burst}N_{cpb}),N_{hop}N_{cpb}\right)$. The nonzero values of $\widetilde{R}_{xx}(2\alpha_q^x,\tau)$ are related to $\widetilde{\phi}_p(\alpha_q^x,\tau)$. Moreover, the range of the nonzero support of $\widetilde{R}_{xx}(2\alpha_q^x,\tau)$ w.r.t. $\tau$ increases, since $\widetilde{p}(t)$ may contain many multipath components. For notation simplicity, we do not consider channel in the detector design, but we show the channel effect in the simulation.
\end{rema}

\section{UWB Signal Detection Using Its Cyclic Features}\label{sec:CF_det}
According to the analysis in the previous section, the CAF of the UWB signal is nonzero at several CFs and for a range of TLs. Hence, multi-cycle multi-lag detectors can be exploited to take full advantage of their cyclic features. Moreover, both the CAF and the conjugate CAF can be used here, as both of them indicate the cyclic features. In this section, we propose constant false alarm rate (CFAR) detectors based on the estimated CAFs to tradeoff between computational complexity and detection performance. Since our proposed detectors are composed of the single-cycle single-lag Dandawate-Giannakis test statistics, we would first briefly review the general form of the Dandawate-Giannakis detector in \cite{Dandawate1994,Lunden2009} in Section~\ref{subsec:DG}. Consequentially, the CFAR detectors are proposed in Section~\ref{subsec:D_pro}. In order to facilitate fair comparisons, four kinds of existing detectors in \cite{Lunden2009,Huang2013} are summarized in Section~\ref{subsec:Others}.

The estimated CAFs at the interested TLs and CFs are used to calculate the test statistics. Hence, they are collected in a row vector $\hat{\br}_{xx^{\left({}^{}_{\ast}\right)}}$ as
\beqa
\hat{\br}_{xx^{\left({}^{}_{\ast}\right)}} = \left[\hat{\br}_{xx^{\left({}^{}_{\ast}\right)}}^1\, \dots\, \hat{\br}_{xx^{\left({}^{}_{\ast}\right)}}^M\right],
\enqa where $M$ is the number of the CFs of interest, the row vector $\hat{\br}_{xx^{\left({}^{}_{\ast}\right)}}^i$ is defined as
\beqa
\hat{\br}_{xx^{\left({}^{}_{\ast}\right)}}^i \!=\!\left[{\rm Re}\left\{\hat{R}_{xx^{\left({}^{}_{\ast}\right)}}(\alpha_i,\tau_{i,1})\right\}, \dots, {\rm Re}\left\{\hat{R}_{xx^{\left({}^{}_{\ast}\right)}}(\alpha_i,\tau_{i,N_i})\right\}, \right.\nonumber\\
\!\!\left.{\rm Im}\left\{\hat{R}_{xx^{\left({}^{}_{\ast}\right)}}(\alpha_i,\tau_{i,1})\right\}, \dots, {\rm Im}\left\{\hat{R}_{xx^{\left({}^{}_{\ast}\right)}}(\alpha_i,\tau_{i,N_i})\right\}\right],\nonumber
\enqa and
\beqa
\hat{R}_{xx^{\left({}^{}_{\ast}\right)}}(\alpha_i,\tau_{i,l})\!=\!\frac{1}{K}\sum_{n=0}^{K-1}x[n]x^{\left({}_{}^{\ast}\right)}[n + \tau_{i,l}]e^{-j2\pi\alpha_i n}, \\
\!\! \forall i \in \{ 1,\dots,M\}, \,\, \forall l \in \{ 1,\dots,N_i\},\nonumber
\enqa with $K$ being the number of samples of the UWB signal ($x(t)$), $\alpha_i$ being the CF of interest, $\tau_{i,l}$ being the $l$th TL of interest for $\alpha_i$, and $N_{i}$ being the total number of TLs for $\alpha_i$. Thus, $2J = 2\sum_{i=1}^{M}N_{i}$ is the total length of $\hat{\br}_{xx^{\left({}^{}_{\ast}\right)}}$. Note that $x[n]$'s are the discrete samples of $x(t)$.



The detection of the UWB signal is a binary-hypothesis test. The two hypotheses are given as follows:
\beqa\label{BHT}
\begin{array}{ccl}
\mathcal{H}_0&:& \hat{\br}_{xx^{\left({}^{}_{\ast}\right)}} = \mathbf{e}, \\
\mathcal{H}_1&:& \hat{\br}_{xx^{\left({}^{}_{\ast}\right)}} = {\br}_{xx^{\left({}^{}_{\ast}\right)}} + \mathbf{e},\end{array}
\enqa where ${\br}_{xx^{\left({}^{}_{\ast}\right)}}$ is the true nonrandom CAF vector, and $\mathbf{e}$ is the estimation error row vector, which is asymptotically zero-mean Gaussian distributed with covariance matrix $\sigma^2\mathbf{I}$. For the above binary-hypothesis test, several existing detectors are reviewed for comparison, and low-complexity CFAR detectors are proposed in the following subsections.

\subsection{The multi-cycle multi-lag Dandawate-Giannakis detector}\label{subsec:DG}

For the binary-hypotheses test (\ref{BHT}), the multi-cycle multi-lag detector \cite{Lunden2009}, which is a natural extension of the original Dandawate-Giannakis detector \cite{Dandawate1994}, is given by
\beqa\label{Dandawate_detector}
\mathcal{T}_{{\rm DG}^{\left({}^{}_{\ast}\right)}} = K \hat{\br}_{xx^{\left({}^{}_{\ast}\right)}} \hat{\mbox{\boldmath$\Sigma$}}^{-1}_{xx^{\left({}^{}_{\ast}\right)}} \left(\hat{\br}_{xx^{\left({}^{}_{\ast}\right)}}\right)^T,
\enqa where $(\cdot)^T$ denotes the transpose, and $\hat{\mbox{\boldmath$\Sigma$}}_{xx^{\left({}^{}_{\ast}\right)}}$ is the estimated asymptotic covariance matrix following the method in \cite{Dandawate1994,Lunden2009}. The true covariance matrix ${\mbox{\boldmath$\Sigma$}}_{xx^{\left({}^{}_{\ast}\right)}}$ can be divided into $M^2$ blocks of size $2N_i \times 2N_{\ell},\,\, \forall i , \forall \ell \in \mathcal{M}$, where $\mathcal{M}=\{ 1,\dots,M\}$. The block ${\mbox{\boldmath$\Sigma$}}_{xx^{\left({}^{}_{\ast}\right)}}(i, \ell)$ of size $2N_i \times 2N_{\ell}$ is the covariance matrix for the CF pair $(\alpha_i, \alpha_\ell)$. Thus, it is given by \cite{Dandawate1994,Lunden2009}
\beqa\label{cov_est}
{\mbox{\boldmath$\Sigma$}}_{xx^{\left({}^{}_{\ast}\right)}}(i,\ell)\!\!= \! \!\left[\begin{array}{cc}
                                              \!{\rm Re}\left\{\displaystyle\frac{\bQ_{i,\ell}^{{\left({}^{}_{\ast}\right)}} + \bP_{i,\ell}^{{\left({}^{}_{\ast}\right)}}}{2}\right\}\! &\! {\rm Im}\left\{\displaystyle\frac{\bQ_{i,\ell}^{{\left({}^{}_{\ast}\right)}} - \bP_{i,\ell}^{{\left({}^{}_{\ast}\right)}}}{2}\right\} \\
                                              \quad\\
\!{\rm Im}\left\{\displaystyle\frac{\bQ_{i,\ell}^{{\left({}^{}_{\ast}\right)}} + \bP_{i,\ell}^{{\left({}^{}_{\ast}\right)}}}{2}\right\}\!&\! {\rm Re}\left\{\displaystyle\frac{\bP_{i,\ell}^{{\left({}^{}_{\ast}\right)}} - \bQ_{i,\ell}^{{\left({}^{}_{\ast}\right)}}}{2}\right\}
                                            \end{array}
\right],\nonumber
\enqa where $\bQ_{i,\ell}^{{\left({}^{}_{\ast}\right)}}$ and $\bP_{i,\ell}^{{\left({}^{}_{\ast}\right)}}$ are defined as
\beqa
\bQ_{i,\ell}^{{\left({}^{}_{\ast}\right)}}(k,l) &=& S^{{\left({}^{}_{\ast}\right)}}_{\tau_{\ell,k} \tau_{i,l}} (\alpha_i+\alpha_\ell,\alpha_\ell),\\
\bP_{i,\ell}^{{\left({}^{}_{\ast}\right)}}(k,l) &=& S^{{\left({}^{}_{\ast}\right)},\ast}_{\tau_{\ell,k}\tau_{i,l}}(\alpha_i-\alpha_\ell,-\alpha_\ell),
\enqa and the estimates of $S^{{\left({}^{}_{\ast}\right)}}_{\tau_{\ell,k} \tau_{i,l}} (\alpha_i+\alpha_\ell,\alpha_\ell)$ and $S^{{\left({}^{}_{\ast}\right)},\ast}_{\tau_{\ell,k}\tau_{i,l}}(\alpha_i-\alpha_\ell,-\alpha_\ell)$ are calculated using the frequency smoothed cyclic periodograms, respectively
\beqa
\label{SCD1}&&\hat{S}^{\left({}^{}_{\ast}\right)}_{\tau_{\ell,k} \tau_{i,l}} (\alpha_i+\alpha_\ell,\alpha_\ell)\\
\!\!&\!\!\!=\!\!\!&\!\!\frac{1}{KL}\!\!\sum_{s=-(L-1)/2}^{(L-1)/2}\!\!W(s)
F^{\left({}^{}_{\ast}\right)}_{\tau_{i,l}}\left(\alpha_i - \frac{ s}{K}\right)F^{\left({}^{}_{\ast}\right)}_{\tau_{\ell,k}}\left(\alpha_\ell + \frac{ s}{K}\right),\nonumber\\
\label{SCD2}&&\hat{S}^{{\left({}^{}_{\ast}\right)},\ast}_{\tau_{j,m}\tau_{i,n}}(\alpha_i-\alpha_\ell,-\alpha_\ell)\\
\!\!&\!\!\!=\!\!\!&\!\!\frac{1}{KL}\!\!\sum_{s=-(L-1)/2}^{(L-1)/2}\!\!W(s)
\left(F^{\left({}^{}_{\ast}\right)}_{\tau_{i,l}}\left(\alpha_i + \frac{ s}{K}\right)\right)^\ast F^{\left({}^{}_{\ast}\right)}_{\tau_{\ell,k}}\left(\alpha_\ell + \frac{ s}{K}\right),\nonumber
\enqa where $F^{\left({}^{}_{\ast}\right)}_{\tau}(\alpha)=\sum_{n=0}^{K-1}x[n]x^{\left({}_{}^{\ast}\right)}[n+\tau]e^{-j2\pi \alpha n}$, and $W(s)$ is the normalized spectral smoothing window function with length $L$. Under the null hypothesis $\mathcal{H}_0$, the distribution of $\mathcal{T}_{{\rm DG}^{\left({}^{}_{\ast}\right)}}$ asymptotically converges to the central $\chi_{2J}^2$ distribution with $2J$ degrees of freedom. Therefore, a threshold $\gamma_{\rm DG}$ can be decided by a constant false alarm rate as $P_{\rm fa} = {\rm Prob}(\mathcal{T}_{{\rm DG}^{\left({}^{}_{\ast}\right)}} \ge \gamma_{\rm DG}|\mathcal{H}_0)$.

\subsection{Proposed CFAR detectors}\label{subsec:D_pro}
The computational complexity of the detector (\ref{Dandawate_detector}) including (\ref{SCD1}) and (\ref{SCD2}) is notably high. To reduce the complexity, we propose a computationally efficient cyclic detector as
\beqa\label{T_h_1}
\mathcal{T}_{{\rm prop}^{\left({}^{}_{\ast}\right)}}^{\rm I} = \sum_{i = 1}^{M}\mathcal{Y}^{\rm I}_{i^{\left({}^{}_{\ast}\right)}} ,
\enqa
where
\beqa
\mathcal{Y}^{\rm I}_{i^{\left({}^{}_{\ast}\right)}} = \max_{l=1,\dots,N_i}\mathcal{T}_{{\rm DG}^{\left({}^{}_{\ast}\right)}}(\alpha_i,\tau_{i,l}).
\enqa The test statistic $\mathcal{T}_{{\rm DG}^{\left({}^{}_{\ast}\right)}}(\alpha_i,\tau_{i,l})$ is a single-cycle single-lag Dandawate-Giannakis detector, and thus a special case of
(\ref{Dandawate_detector}). It is given by
\beqa
\mathcal{T}_{{\rm DG}^{\left({}^{}_{\ast}\right)}}(\alpha_i,\tau_{i,l})= K \hat{\br}_{xx^{\left({}^{}_{\ast}\right)}}^{i,l} \hat{\mbox{\boldmath$\Sigma$}}^{-1}_{xx^{\left({}^{}_{\ast}\right)}}(i,i,l,l) \left(\hat{\br}_{xx^{\left({}^{}_{\ast}\right)}}^{i,l}\right)^T,
\enqa where $\hat{\br}_{xx^{\left({}^{}_{\ast}\right)}}^{i,l} \!=\! \left[ {\rm Re}\left\{\hat{R}_{xx^{\left({}^{}_{\ast}\right)}}(\alpha_i,\tau_{i,l})\right\}, {\rm Im}\left\{\hat{R}_{xx^{\left({}^{}_{\ast}\right)}}(\alpha_i,\tau_{i,l})\right\}\right]$ and
\beqa\label{cov_est}
{\mbox{\boldmath$\Sigma$}}_{xx^{\left({}^{}_{\ast}\right)}}(i,i,l,l)\quad\quad\quad\quad\quad\quad\quad\quad \quad\quad\quad\quad\quad\quad\quad\quad\quad\quad\quad\quad\nonumber\\
\!\!= \! \!\left[\begin{array}{cc}
                                              \!\!{\rm Re}\left\{\displaystyle\frac{\bQ_{i,i}^{{\left({}^{}_{\ast}\right)}}(l,l) + \bP_{i,i}^{{\left({}^{}_{\ast}\right)}}(l,l)}{2}\right\}\! &\!\! {\rm Im}\left\{\displaystyle\frac{\bQ_{i,i}^{{\left({}^{}_{\ast}\right)}}(l,l) - \bP_{i,i}^{{\left({}^{}_{\ast}\right)}}(l,l)}{2}\right\} \\
                                              \quad\\
\!\!{\rm Im}\left\{\displaystyle\frac{\bQ_{i,i}^{{\left({}^{}_{\ast}\right)}}(l,l) + \bP_{i,i}^{{\left({}^{}_{\ast}\right)}}(l,l)}{2}\right\}\!&\!\! {\rm Re}\left\{\displaystyle\frac{\bP_{i,i}^{{\left({}^{}_{\ast}\right)}}(l,l) - \bQ_{i,i}^{{\left({}^{}_{\ast}\right)}}(l,l)}{2}\right\}
                                            \end{array}
\!\!\!\!\right].\nonumber
\enqa


Furthermore, the mapping between the CF set and the TL set can also be described in another way as follows: for each TL $\tau_{u}, u = 1,\dots, \widetilde{N}$ with $\widetilde{N}$ being the total number of TLs of interest, the CFs of interest are $\alpha_{u,v}, v = 1,\dots,\widetilde{M}_u$, where $\widetilde{M}_u$ is the total number of CFs for the TL $\tau_u$. Note that $2J = 2\sum_{u = 1}^{\widetilde{N}}\widetilde{M}_u= 2\sum_{i = 1}^{M}N_i$ is the total length of $\hat{\bf{r}}_{xx^{\left({}^{}_{\ast}\right)}}$. Hence, a nature variation of the test statistic $\mathcal{T}^{\rm I}_{{\rm prop}^{\left({}^{}_{\ast}\right)}}$ in (\ref{T_h_1}) is proposed as
\beqa\label{T_h_2}
\mathcal{T}_{{\rm prop}^{\left({}^{}_{\ast}\right)}}^{\rm II} = \sum_{u = 1}^{\widetilde{N}}\mathcal{Y}^{\rm II}_{u^{\left({}^{}_{\ast}\right)}} ,
\enqa where
\beqa
\mathcal{Y}^{\rm II}_{u^{\left({}^{}_{\ast}\right)}} = \max_{v=1,\dots,\widetilde{M}_u}\mathcal{T}_{{\rm DG}^{\left({}^{}_{\ast}\right)}}(\alpha_{u,v},\tau_{u}).
\enqa Both $\mathcal{T}_{{\rm prop}^{\left({}^{}_{\ast}\right)}}^{\rm I}$ (\ref{T_h_1}) and $\mathcal{T}_{{\rm prop}^{\left({}^{}_{\ast}\right)}}^{\rm II}$ (\ref{T_h_2}) are motivated by the rich cyclic features of the UWB signal, since multiple CFs and TLs could provide the diversity gain to combat the multipath fading.


As the methods to calculate the thresholds for $\mathcal{T}_{{\rm prop}^{\left({}^{}_{\ast}\right)}}^{\rm I}$ and $\mathcal{T}_{{\rm prop}^{\left({}^{}_{\ast}\right)}}^{\rm II}$ are the same, the calculation of the threshold for $\mathcal{T}_{{\rm prop}^{\left({}^{}_{\ast}\right)}}^{\rm I}$ is exemplified. Under the null hypothesis $\mathcal{H}_0$, it is known that $\mathcal{T}_{{\rm DG}^{\left({}^{}_{\ast}\right)}}(\alpha_i,\tau_{i,l})$ asymptotically follows the central $\chi_{2}^2$ distribution. The probability of density function (pdf) $p_i(y)$ of $\mathcal{Y}_{i^{\left({}^{}_{\ast}\right)}}^{\rm I}$ ($y \triangleq \mathcal{Y}_{i^{\left({}^{}_{\ast}\right)}}^{\rm I}$) can be computed as
\beqa
f_i^{\rm I}(y) = N_iF(y)^{N_i-1}f(y), \,\, \forall i \in \mathcal{M},
\enqa where $F(y)$ and $f(y)$ are the cumulative distribution function (cdf) and the pdf of the central $\chi_{2}^2$ given by
\beqa
F(y) &=& 1 - e^{-\frac{y}{2}}, y\ge 0,\\
f(y) &=&\frac{1}{2}e^{-\frac{y}{2}}, y \ge 0.
\enqa
Making use of the binomial expansion, we arrive at
\beqa
f_i^{\rm I}(y)\! =\! \frac{N_i}{2}\sum_{k=0}^{N_i-1}(-1)^k\left(\begin{array}{c}
{N_i-1}\\
{k}
\end{array}\right)e^{-\frac{(k+1)y}{2}},\nonumber\\
 y \ge 0,\,\, \forall i \in \mathcal{M}.
\enqa Since $\mathcal{Y}^{\rm I}_{i^{\left({}^{}_{\ast}\right)}}$ are independent of each other, the pdf of the test statistic of $\mathcal{T}_{{\rm prop}^{\left({}^{}_{\ast}\right)}}^{\rm I}$ should be a multi-dimensional convolution of all $f_i^{\rm I}(y), \forall i \in \mathcal{M}$. Therefore, we can achieve the pdf of $\mathcal{T}_{{\rm prop}^{\left({}^{}_{\ast}\right)}}^{\rm I}$ numerically and compute a threshold $\gamma_{\rm prop}^{\rm I}$ according to a constant false alarm rate as $P_{\rm fa} = {\rm Prob}(\mathcal{T}_{{\rm prop}^{\left({}^{}_{\ast}\right)}}^{\rm I} \ge \gamma_{\rm prop}^{\rm I}|\mathcal{H}_0)$. In the case that $N_i = N_\ell, \forall i,\forall \ell \in \mathcal{M}$ and $M$ is small, it is possible to obtain a closed-form pdf of $\mathcal{T}^{\rm I}_{{\rm prop}^{\left({}^{}_{\ast}\right)}}$. For example, when $M = 2$ and $N_1 = N_2 = 2$, the pdf and the cdf of $\mathcal{T}_{{\rm prop}^{\left({}^{}_{\ast}\right)}}^{\rm I}$ (Let us reload $y$ as $y \triangleq \mathcal{T}^{\rm I}_{{\rm prop}^{\left({}^{}_{\ast}\right)}}$) are given by, respectively
\beqa
\widetilde{f}(y) = (4+y)(e^{-y} + e^{-y/2}),
\enqa and
\beqa
\widetilde{F}(y) = 1+(4-2y)e^{-y/2} -(5+y)e^{-y}.
\enqa As a result, the threshold $\gamma_{\rm prop}^{\rm I}$ can be found in a lookup table calculated offline in advance.

\subsection{Other existing detectors}\label{subsec:Others}
There are several variations of the multi-cycle multi-lag Dandawate-Giannakis detector $\mathcal{T}_{{\rm DG}^{\left({}^{}_{\ast}\right)}}$ in (\ref{Dandawate_detector}). Regardless of the correlation between different CFs, the estimated covariance matrix can be simplified as a block diagram matrix. Denote the corresponding test statistic as $\mathcal{T}_{{\rm sum\_DG}^{\left({}^{}_{\ast}\right)}}$, which is given by \beqa\label{T_s}
\mathcal{T}_{{\rm sum\_DG}^{\left({}^{}_{\ast}\right)}} = \sum_{i=1}^{M}\mathcal{T}_{{\rm DG}^{\left({}^{}_{\ast}\right)}}(\alpha_i),
\enqa where
\beqa
\mathcal{T}_{{\rm DG}^{\left({}^{}_{\ast}\right)}}(\alpha_i) = K \hat{\br}_{xx^{\left({}^{}_{\ast}\right)}}^{i} \hat{\mbox{\boldmath$\Sigma$}}^{-1}_{xx^{\left({}^{}_{\ast}\right)}}(i,i) (\hat{\br}^{i}_{xx^{\left({}^{}_{\ast}\right)}})^T.
\enqa Note that $\mathcal{T}_{{\rm DG}^{\left({}^{}_{\ast}\right)}}(\alpha_i)$ is the test statistic for a single CF with multiple TLs. When only a single TL is employed for each CF, the test statistic $\mathcal{T}_{{\rm sum\_DG}^{\left({}^{}_{\ast}\right)}}$ is equivalent to $\mathcal{T}_{{\rm prop}^{\left({}^{}_{\ast}\right)}}^{\rm I}$. Another variation is to choose the maximum test statistic among $\mathcal{T}_{{\rm DG}^{\left({}^{}_{\ast}\right)}}(\alpha_i), i \in \mathcal{M}$, and compare it with a threshold. Let us denote this test statistic as
\beqa\label{T_m}
\mathcal{T}_{{\rm max\_DG}^{\left({}^{}_{\ast}\right)}} = \max_{i=1,\dots,M}\mathcal{T}_{{\rm DG}^{\left({}^{}_{\ast}\right)}}(\alpha_i).
\enqa The test statistics $\mathcal{T}_{{\rm sum\_DG}^{\left({}^{}_{\ast}\right)}}$ and $\mathcal{T}_{{\rm max\_DG}^{\left({}^{}_{\ast}\right)}}$ are more computationally efficient than $\mathcal{T}_{{\rm DG}^{\left({}^{}_{\ast}\right)}}$. Furthermore, it has been shown in \cite{Lunden2009} that the detection performance of $\mathcal{T}_{{\rm sum\_DG}^{\left({}^{}_{\ast}\right)}}$ is close to $\mathcal{T}_{{\rm DG}^{\left({}^{}_{\ast}\right)}}$.


Moreover, an ad hoc detector $\mathcal{T}_{{\rm ad\_hoc}}$ is proposed in \cite{Huang2013} to take colored Gaussian noise into account for the null hypothesis. In order to facilitate a fair comparison in the simulation section, we review the detector proposed in \cite{Huang2013} using the notations defined in this paper. The autocorrelation of the colored Gaussian noise is assumed to be nonzero in the range of $[-L_n, L_n]$ ($\widetilde{R}_{xx}(m) \neq 0, |m| \le L_n$). Thus, the test statistic based on the CAF estimates is derived as
\beqa\label{T_c}
\mathcal{T}_{{\rm ad\_hoc}} = 2K\sum_{i=1}^{M}\sum_{l=1}^{N_i} \frac{|\hat{R}_{xx}(\alpha_i , \tau_{i,l})|^2}{\hat{\gamma}_{xx}(\alpha_i)},
\enqa where
\beqa
\hat{\gamma}_{xx}(\alpha_i) &=& \sum_{s=-L_n}^{L_n}|\hat{\widetilde{R}}_{xx}(s)|^2e^{-j2\pi\alpha_i s},\\
\hat{\widetilde{R}}_{xx}(s) &=& \frac{1}{K-s}\sum_{n=0}^{K-s-1}x[n]x^\ast[n+s], \, s \ge 0,
\enqa and $\hat{\widetilde{R}}_{xx}(-s) = \hat{\widetilde{R}}_{xx}^\ast(s)$. On the other hand, the test statistic based on the conjugate CAF estimates is adapted as
\beqa\label{T_c1}
\mathcal{T}_{{\rm ad\_hoc}^{\ast}} = 2K\sum_{i=1}^{M}\sum_{l=1}^{N_i} \frac{|\hat{R}_{xx^\ast}(\alpha_i , \tau_{i,l})|^2}{\hat{\gamma}_{xx^\ast}(\alpha_i,\tau_{i,l})},
\enqa where
\beqa
&&\hat{\gamma}_{xx^\ast}(\alpha_i,\tau_{i,l})\\
&\!=\!&\!\!\sum_{s=-L_n}^{L_n}\!\!\!\left(\hat{\widetilde{R}}_{xx}^2(s) + \hat{\widetilde{R}}_{xx}(s + \tau_{i,l})\hat{\widetilde{R}}_{xx}(s-\tau_{i,l})\right)e^{j2\pi\alpha_i s}.\nonumber
\enqa  The correlation among the CAF estimates with different CF-TL pairs are neglected in (\ref{T_c}) and (\ref{T_c1}). Moreover, the test statistic $\mathcal{T}_{{\rm ad\_hoc}^{\left({}^{}_{\ast}\right)}}$ asymptotically follows a central $\chi_{2J}^2$ distribution under the null hypothesis.



Last but not the least, the energy detector is widely used in the signal detection due to its simplicity. The test statistic for energy detector is given by
\beqa\label{T_e}
\mathcal{T}_{\rm ED} = \frac{1}{K}\sum_{n=0}^{K-1}|x[n]|^2,
\enqa where
\beqa
\mathcal{H}_0: \mathcal{T}_{\rm ED} \sim \mathcal{N}\left(\varsigma_n^2, \frac{2\varsigma_n^4}{K}\right),
\enqa according to the central limit theorem, and $\varsigma_n^2$ is the variance of the observation noise.

\section{The computational complexity analysis}\label{sec:complx}

In this section, we analyze the computational complexity of the proposed detectors, and compare them with other detectors reviewed in Section~\ref{sec:CF_det}. The computational complexity counts for the number of operations to calculate the test statistics. The complexities of additions, subtractions and comparison are trivial compared to multiplications and divisions, and thus they are neglected for simplicity. We further assume that the complexities of the multiplication and the division are the same, thus the divisions are counted as the multiplications as well. The complexities of the elements to calculate the test statistics is first explored. The computational complexity to calculate $\hat{{R}}_{xx^{\left({}^{}_{\ast}\right)}}(\alpha_i,\tau_{i,l})$ is $2K$ multiplications. The calculation of $\hat{{\mbox{\boldmath$\Sigma$}}}_{xx^{\left({}^{}_{\ast}\right)}}(i,\ell)$ involves $N_iN_{\ell}(6K+4L)$ multiplications. Hence, to achieve $\hat{{\mbox{\boldmath$\Sigma$}}}_{xx^{\left({}^{}_{\ast}\right)}}$ needs $(6K+4L)(\sum_{i=1}^MN_i^2 + \sum_{i=1}^{M-1}\sum_{\ell=i+1}^{M}N_iN_\ell)$ multiplications. The inverse of an $A \times A$ matrix using the Gauss-Jordan elimination requires $A^3+6A^2$ multiplications. As a result, the inverse of $\hat{{\mbox{\boldmath$\Sigma$}}}_{xx^{\left({}^{}_{\ast}\right)}}$ and $\hat{{\mbox{\boldmath$\Sigma$}}}_{xx^{\left({}^{}_{\ast}\right)}}(i,i)$ count for $8J^3 + 24J^2$ and $8N_i^3 + 24N_i^2$ multiplications, respectively. Moreover, the product of an $A \times B$ matrix and a $B \times C$ matrix needs $ABC$ multiplications. Therefore, the test statistic $\mathcal{T}_{{\rm DG}^{\left({}^{}_{\ast}\right)}}(\alpha_i,\tau_{i,l})$ itself for a single $(\alpha_i,\tau_{i,l})$ pair asks for $6$ multiplications, and the calculation of $\mathcal{T}_{{\rm DG}^{\left({}^{}_{\ast}\right)}}(\alpha_i)$ needs $4N_i^2 + 2N_i$ multiplications. To compute the test statistic $\mathcal{T}_{{\rm DG}^{\left({}^{}_{\ast}\right)}}$ requires $4J^2+2J$ multiplications. Considering the test statistics $\mathcal{T}_{{\rm ad\_hoc}^{\left({}^{}_{\ast}\right)}}$, the computation of the correlation term $\hat{\widetilde{R}}_{xx}(s)$ involves approximately $K$ multiplications. There are $L_n+1$ TLs. Thus, the correlation coefficients $\hat{\gamma}_{xx}(\alpha_i)$ and $\hat{\gamma}_{xx}(\alpha_i,\tau_{i,\ell})$ need $2(L_n + 1)$ and $3(L_n+1)$ multiplications, respectively. The test statistic $\mathcal{T}_{{\rm ad\_hoc}^{\left({}^{}_{\ast}\right)}}$ itself asks for $2J$ multiplications.

\begin{table*}
\centering
\begin{tabular}{|c|c|c|c|}
  \hline
  Detectors & $\hat{{\mbox{\boldmath$\Sigma$}}}_{xx^{\left({}^{}_{\ast}\right)}}(i,i,l,l)$/$\hat{{\mbox{\boldmath$\Sigma$}}}_{xx^{\left({}^{}_{\ast}\right)}}(i,\ell)$ / $\hat{{\mbox{\boldmath$\Sigma$}}}_{xx^{\left({}^{}_{\ast}\right)}}$ & Matrix inverse &$\mathcal{T}_{{\rm DG}^{\left({}^{}_{\ast}\right)}}(\alpha_i,\tau_{i,l})$ / $\mathcal{T}_{{\rm DG}^{\left({}^{}_{\ast}\right)}}(\alpha_i)$\\
  \hline
  $\mathcal{T}^{\rm I}_{{\rm prop}^{\left({}^{}_{\ast}\right)}}$, $\mathcal{T}^{\rm II}_{{\rm prop}^{\left({}^{}_{\ast}\right)}}$ & $J(6K+4L)$ & $32J$  & $6J$\\
  \hline
  $\mathcal{T}_{{\rm DG}^{\left({}^{}_{\ast}\right)}}$   & $\displaystyle(6K+4L)\left(\sum_{i=1}^MN_i^2 + \sum_{i=1}^{M-1}\sum_{\ell=i+1}^{M}N_iN_\ell\right)$ & $8J^3 + 24J^2$ & -\\
  \hline
  $\mathcal{T}_{{\rm sum\_DG}^{\left({}^{}_{\ast}\right)}}$, $\mathcal{T}_{{\rm max\_DG}^{\left({}^{}_{\ast}\right)}}$ & $\displaystyle(6K+4L)\sum_{i=1}^M N_i^2$ &$\displaystyle\sum_{i=1}^M \left(8N_i^3 + 24N_i^2\right)$& $\displaystyle\sum_{i=1}^M \left(4N_i^2 + 2N_i\right)$\\
  \hline
  \hline
  Detectors & \multicolumn{3}{|c|}{Total number of operations}\\
  \hline
  $\mathcal{T}^{\rm I}_{{\rm prop}^{\left({}^{}_{\ast}\right)}}$, $\mathcal{T}^{\rm II}_{{\rm prop}^{\left({}^{}_{\ast}\right)}}$ &\multicolumn{3}{|c|}{ $2JK + J(6K+4L) + 38J$}\\
  \hline
  $\mathcal{T}_{{\rm DG}^{\left({}^{}_{\ast}\right)}}$ &\multicolumn{3}{|c|}{$\displaystyle 2JK + (6K+4L)\left(\sum_{i=1}^MN_i^2 + \sum_{i=1}^{M-1}\sum_{\ell=i+1}^{M}N_iN_\ell\right)+ 8J^3 + 28J^2 + 2J$}\\
  \hline
  $\mathcal{T}_{{\rm sum\_DG}^{\left({}^{}_{\ast}\right)}}$, $\mathcal{T}_{{\rm max\_DG}^{\left({}^{}_{\ast}\right)}}$ &\multicolumn{3}{|c|}{$\displaystyle 2JK + \sum_{i=1}^M \left(N_i^2(6K+4L) + 8N_i^3 + 28N_i^2 + 2N_i \right)$}\\
  \hline
  $\mathcal{T}_{{\rm ad\_hoc}^{\left({}^{}_{\ast}\right)}}$ &\multicolumn{3}{|c|}{ $2JK +(L_n+1)K + 2M(L_n+1) + 2J$}\\
  \hline
\end{tabular}
\caption{The computational complexity of different detectors}\label{CC_detector}
\end{table*}

Based on the above analysis, the results are summarized in Table~\ref{CC_detector}. Recall that $J = \sum_{i=1}^MN_i = \sum_{u=1}^{\widetilde{N}}\widetilde{M}_u$ and $K$ is the number of samples. Therefore $K \gg J, L, L_n$. The terms related to $K$ in Table~\ref{CC_detector} are the most significant ones. The proposed detectors $\mathcal{T}^{\rm I}_{{\rm prop}^{\left({}^{}_{\ast}\right)}}$ and $\mathcal{T}^{\rm II}_{{\rm prop}^{\left({}^{}_{\ast}\right)}}$ have comparable complexities as the detector $\mathcal{T}_{{\rm ad\_hoc}^{\left({}^{}_{\ast}\right)}}$. All of them need much less operations than the detector $\mathcal{T}_{{\rm DG}^{\left({}^{}_{\ast}\right)}}$ and its variations ($\mathcal{T}_{{\rm sum\_DG}^{\left({}^{}_{\ast}\right)}}$ and $\mathcal{T}_{{\rm max\_DG}^{\left({}^{}_{\ast}\right)}}$). The detector $\mathcal{T}_{{\rm DG}^{\left({}^{}_{\ast}\right)}}$ requires the most computational resources.

\section{Numerical Examples}\label{sec:sim}
In this section, the performance of the proposed detectors are evaluated under several scenarios by Monte-Carlo simulations. An $8$th order Butterworth pulse with a $3\,{\rm dB}$ bandwidth of $500\,{\rm MHz}$ \cite{S802.15.4a} is used as the baseband UWB pulse. The IEEE 802.15.4a multipath channel model CM1 - indoor LOS \cite{Chan_802.15.4a} is employed for simulations. The channel changes randomly in each Monte-Carlo run. The average power of the channels are normalized. Thus, the received baseband UWB signal is a complex signal due to the multipath channel effect. Furthermore, in order to avoid the cyclic spectrum aliasing, the sampling rate is set to $1\,{\rm GHz}$ in the simulations. Without loss of generality, we are interested in detecting the UWB signal with the highest symbol rate ($31.2\,{\rm MHz}$). The parameters are set as $N_{burst} = 8$, $N_{hop} = 2$, $N_{cpb} = 2$ and $T_c = 2\,{\rm ns}$ \cite{S802.15.4a}. Therefore, the fundamental cyclic frequency of the UWB signal is that $2\alpha_1^x = 2/T_{dsym} = 62.4\,\textrm{MHz}$ according to the derivations in Section~\ref{sec:CF}. The spectrum smoothing window is selected as the Kaiser window of length $65$ with $\beta$ being $1$ to avoid a higher $P_{\rm fa}$ than the expectation based on the remarks in \cite{Huang2013}. The length of the observation window ($T_o$) is $10\,\mu{\rm s}$. Thus, the total number of samples is about $10,000$ ($K = 10,000$). The signal-to-noise ratio (SNR) is defined as $\textrm{SNR} = 10{\rm log}_{10}\sigma_x^2/\sigma_n^2$ similarly as \cite{Lunden2009}, where $\sigma_x^2$ and $\sigma_n^2$ are the power of the UWB signal and the complex Gaussian noise, respectively. The plotted detection performance curves are averaged over $1,000$ Monte-Carlo runs.

\subsection{The cyclic features of the UWB signal}

\begin{figure}
\centering
\centerline{\includegraphics[width=0.75\linewidth]{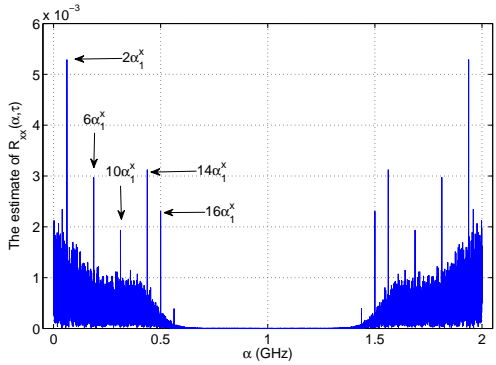}}
\caption{An example of the amplitude of the estimated CAF $\hat{R}_{xx}(\alpha,\tau)$ vs. $\alpha$ without noise and multipath channel, and $\tau = 2\,{\rm ns}$.}\label{CAF_AWGN}
\end{figure}

\begin{figure}
\centering
\centerline{\includegraphics[width=0.75\linewidth]{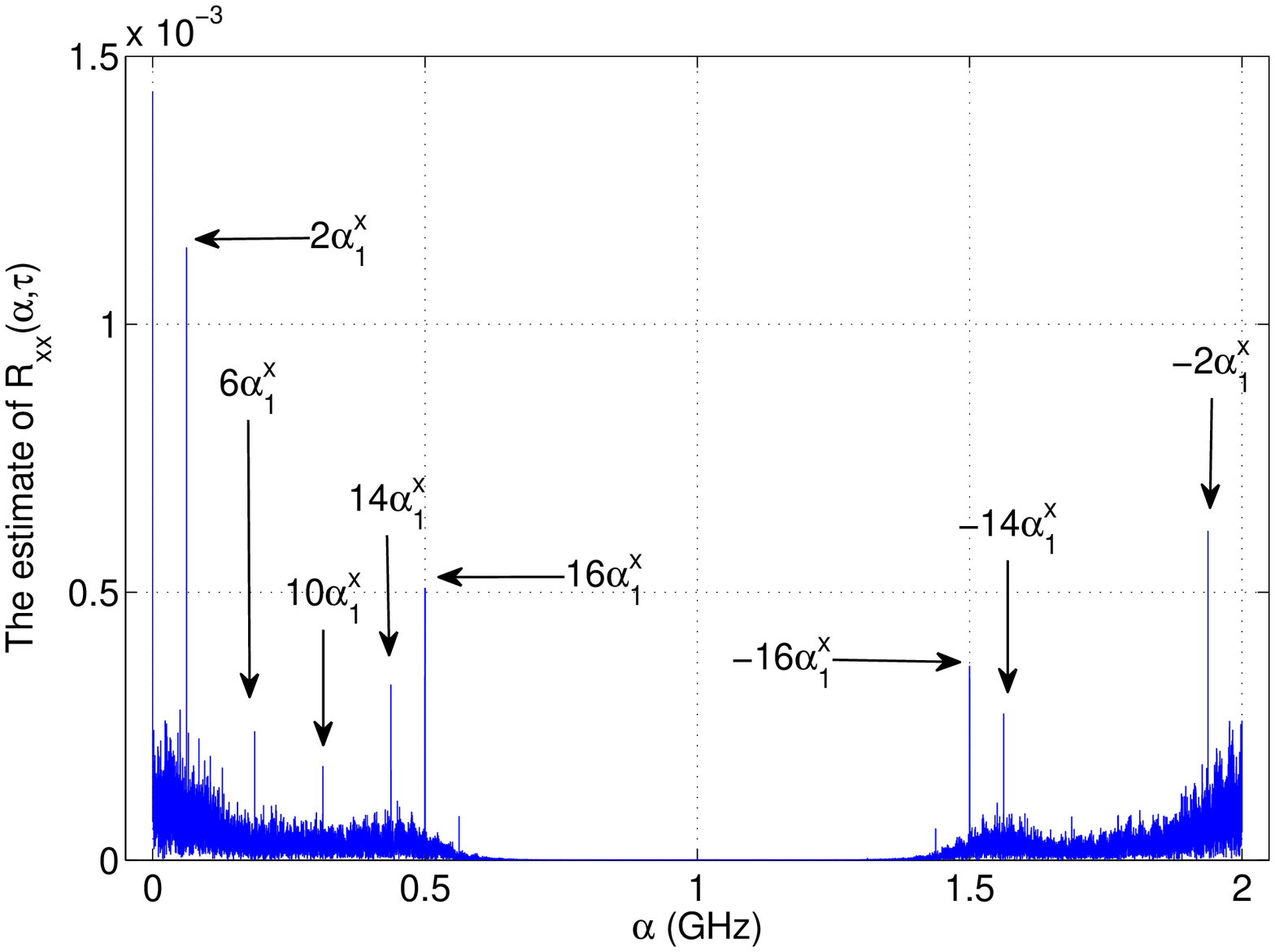}}
\caption{An example of the amplitude of the estimated CAF $\hat{R}_{xx}(\alpha,\tau)$ vs. $\alpha$ under a noiseless multipath channel, and $\tau = 2\,{\rm ns}$.}\label{CAF_FAD_non}
\end{figure}


\begin{figure}
\centering
\centerline{\includegraphics[width=0.75\linewidth]{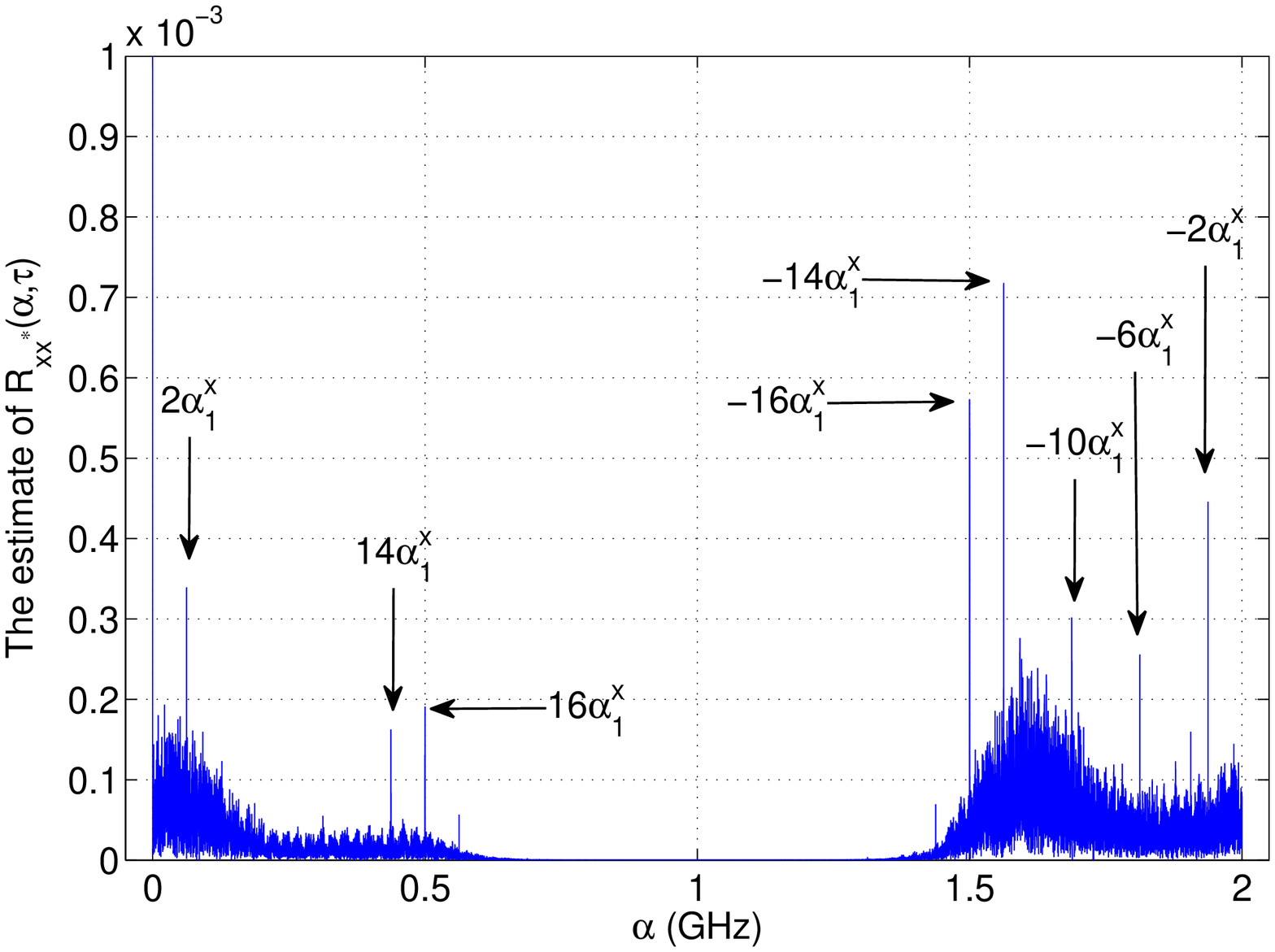}}
\caption{An example of the amplitude of the estimated conjugate CAF $\hat{R}_{xx^\ast}(\alpha,\tau)$ vs. $\alpha$ under a noiseless multipath channel, and $\tau = 2\,{\rm ns}$.}\label{CAF_FAD_conj}
\end{figure}

In order to clearly illustrate the cyclic features of the UWB signal, the sampling rate is $2\,{\rm GHz}$, and the TL is assigned to $2\,{\rm ns}$ in this subsection. Fig.~\ref{CAF_AWGN} shows the amplitude of the estimated CAF $\hat{R}_{xx}(\alpha,\tau)$ of the UWB signal without any noise and multipath effects. As $\hat{R}_{xx}(\alpha,\tau)$ in the range of $[1, 2]\,{\rm GHz}$ is equivalent to the one in the range of $[-1, 0]\,{\rm GHz}$, we denote the estimated CAF $\hat{R}_{xx}(\alpha,\tau)$ in the range of $[1, 2]\,{\rm GHz}$ (or $[0, 1]\,{\rm GHz}$) as $\hat{R}_{xx}(\alpha^{-},\tau)$ (or $\hat{R}_{xx}(\alpha^{+},\tau)$). Without the multipath channel effect, the UWB signal is a real signal. Therefore, $|\hat{R}_{xx}(\alpha^{-},\tau)|$ and $|\hat{R}_{xx}(\alpha^{+},\tau)|$ are symmetric w.r.t. $1$ GHz. Significant peaks of  $|\hat{R}_{xx}(\alpha,\tau)|$ are at the expected CFs $\pm 2\alpha_1^x, \pm 6\alpha_1^x, \pm 10\alpha_1^x, \pm 14\alpha_1^x,\pm 16\alpha_1^x$. The amplitudes of the peaks are in the tendency to decrease as the cycle frequency increases. Since the bandwidth of the UWB signal is $500$ MHz, no significant peaks are observed beyond $500$ MHz.

When the transmitted UWB signal goes through a noiseless multipath channel, the amplitudes of the estimated CAFs of the received UWB signal are depicted in Fig.~\ref{CAF_FAD_non}. Some expected peaks disappear due to the multipath channel effects. Note that $|\hat{R}_{xx^{\left({}_\ast^{}\right)}}(\alpha^{-},\tau)|$ and $|\hat{R}_{xx^{\left({}_\ast^{}\right)}}(\alpha^{+},\tau)|$ are not symmetric anymore. The amplitude of the estimated conjugate CAFs are illustrated in Fig.~\ref{CAF_FAD_conj} as well. Clear peaks at multiple CFs are observed in these figures. Hence, both the CAF and the conjugate CAF estimates can be used to detect the UWB signal.

\subsection{Multi-cycle multi-lag detection under multipath channels}

In this subsection, we investigate the detection performance of various detectors under multipath channels with complex Gaussian noise and colored Gaussian noise, respectively. The false-alarm rate $P_{\rm fa}$ is fixed to $0.01$ for all the detectors.

\subsubsection{Complex Gaussian noise}

\begin{figure}
\centering
\centerline{\includegraphics[width=0.75\linewidth]{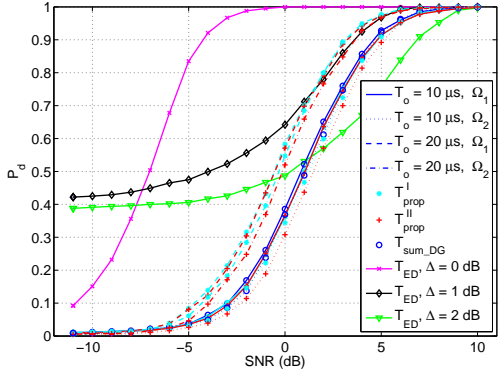}}
\caption{The comparison of $P_{\rm d}$ among different detectors $\mathcal{T}_{\rm prop}^{\rm I}$ (\ref{T_h_1}), $\mathcal{T}_{\rm prop}^{\rm II}$ (\ref{T_h_2}), $\mathcal{T}_{\rm sum\_DG}$ (\ref{T_s}), and $\mathcal{T}_{\rm ED}$ (\ref{T_e}) using different CF and TL sets with different observation window lengths, $P_{\rm fa} = 0.01$.}\label{Fig:Pd_W_To}
\end{figure}



\begin{figure}
\centering
\centerline{\includegraphics[width=0.75\linewidth]{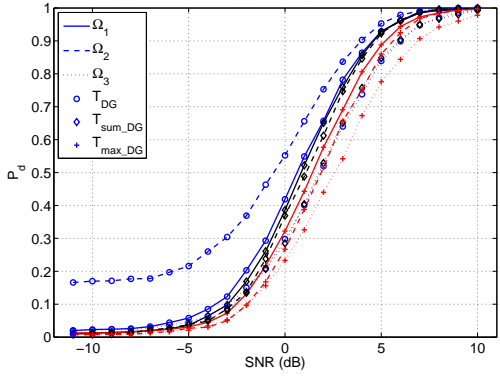}}
\caption{The comparison of $P_{\rm d}$ among different detectors $\mathcal{T}_x$ (\ref{Dandawate_detector}), $\mathcal{T}_{\rm sum\_DG}$ (\ref{T_s}), and $\mathcal{T}_m$ (\ref{T_m}) using different CF and TL sets, $T_o = 10\,\mu{\rm s}$, $P_{\rm fa} = 0.01$.}\label{Fig:Pd_GG_all}
\end{figure}

Two sets of CFs and TLs are chosen. The first set $\Omega_1 = \{\pm 2\alpha_1^x, 2\,{\rm ns}, 4\,{\rm ns}\}$, and the second one $\Omega_2 = \{\pm 2\alpha_1^x, \pm 6\alpha_1^x, 2\,{\rm ns}, 4\,{\rm ns}\}$. Every CF in $\Omega_1$ (or $\Omega_2$) shares the same set of TLs ($\{2\,{\rm ns}, 4\,{\rm ns}\}$). Thus, there are in total four and eight CF-TL pairs for $\Omega_1$ and $\Omega_2$, respectively.


In Fig.~\ref{Fig:Pd_W_To}, the proposed detectors $\mathcal{T}_{\rm prop}^{\rm I}$ and $\mathcal{T}_{\rm prop}^{\rm II}$ are evaluated under multipath channels and with different observation window lengths (denoted by $T_o$ in Fig.~\ref{Fig:Pd_W_To}). Moreover, they are compared with the detectors $\mathcal{T}_{\rm sum\_DG}$ and $\mathcal{T}_{\rm ED}$, which are reviewed in (\ref{T_s}) and (\ref{T_e}), respectively. Witnessed by Fig.~\ref{Fig:Pd_W_To}, employing more CFs for the proposed detectors does not substantially improve the detection performance. On the other hand, due to the increased degree of freedom, the detection performance of the proposed detectors using the parameter set $\Omega_2$ (the dotted and the dash-dot lines) is worse than the one using $\Omega_1$ (the solid and the dashed lines). In general, the $P_{\rm d}$ curves of $\mathcal{T}_{\rm prop}^{\rm I}$ and $\mathcal{T}_{\rm prop}^{\rm II}$ are close to each other. Longer observation time facilities the improvement of the detection. In addition, the detection performance of $\mathcal{T}_{\rm sum\_DG}$ with $T_o = 10 \,{\rm \mu s}$ is depicted in Fig.~\ref{Fig:Pd_W_To} as well. It is slightly better than the performance of the proposed detectors with the same parameter settings. Furthermore, the energy detector $\mathcal{T}_{\rm ED}$ with perfect knowledge of the noise variance ($\Delta = 0$) achieves the best detection performance, when $T_o = 10 {\rm \mu s}$. There is about $8$~dB performance gain of $\mathcal{T}_{\rm ED}$ over the proposed detectors. However, the energy detector is sensitive to noise uncertainty. When the employed noise variance is uniformly distributed between $\pm \Delta \neq 0$ of the perfect noise variance, the detection performance of the energy detector degrades dramatically. Additionally, the energy detector barely guarantees the expected $P_{\rm fa}$ due to the noise uncertainty at low SNR.

In Fig.~\ref{Fig:Pd_GG_all}, the multi-cycle multi-lag Dandawate-Giannakis detector $\mathcal{T}_{\rm DG}$ in (\ref{Dandawate_detector}) is compared with its two variations ($\mathcal{T}_{\rm sum\_DG}$ in (\ref{T_s}) and $\mathcal{T}_{\rm max\_DG}$ in (\ref{T_m})). Besides the sets $\Omega_1$ and $\Omega_2$, another set applied here is that $\Omega_3 = \{\pm 2\alpha_1^x, \pm 6\alpha_1^x, 2\,{\rm ns}\}$. With both $\Omega_1$ and $\Omega_3$, the detection curves of $\mathcal{T}_{\rm sum\_DG}$ closely follow the ones of $\mathcal{T}_{\rm DG}$, while $\mathcal{T}_{\rm max\_DG}$ always performs the worst. In the agreement with Fig.~\ref{Fig:Pd_W_To}, TLs help more than CFs w.r.t. the detection performance. Since the $P_{\rm d}$ of the detector $\mathcal{T}_{\rm DG}$ using the set $\Omega_2$ at low SNR is much higher than the expected $P_{\rm fa}$, the detectors employing the set $\Omega_1$ performs the best overall. Moreover, the detector $\mathcal{T}_{\rm sum\_DG}$ can replace the detector $\mathcal{T}_{\rm DG}$ in the favor of the computational complexity. Hence, the detector $\mathcal{T}_{\rm sum\_DG}$ represents the Dandawate-Giannakis type detectors for comparison in the following subsections.

\subsubsection{Colored Gaussian noise}\label{sec:CN}

\begin{figure}
\centering
\centerline{\includegraphics[width=0.75\linewidth]{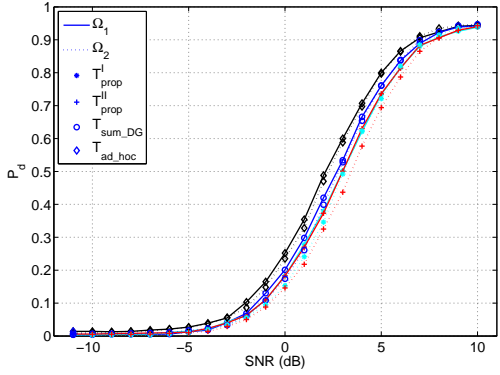}}
\caption{The comparison of $P_{\rm d}$ among different detectors $\mathcal{T}_{\rm prop}^{\rm I}$ (\ref{T_h_1}), $\mathcal{T}_{\rm prop}^{\rm II}$ (\ref{T_h_2}), $\mathcal{T}_{\rm sum\_DG}$ (\ref{T_s}), and $\mathcal{T}_{\rm ad\_hoc}$ (\ref{T_c}) using different CF and TL sets with colored Gaussian noise, $T_o = 10\,\mu{\rm s}$, $P_{\rm fa} = 0.01$.}\label{Pd_Tx_all}
\end{figure}

\begin{figure}
\centering
\centerline{\includegraphics[width=0.75\linewidth]{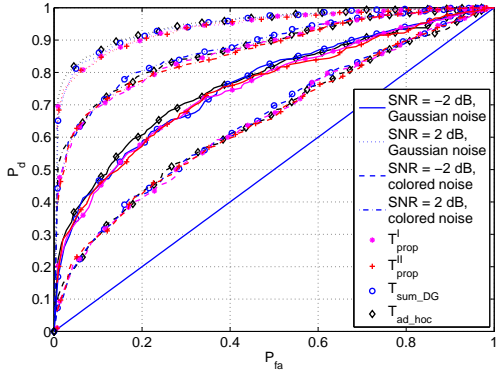}}
\caption{The comparison of ROC among different detectors $\mathcal{T}_{\rm prop}^{\rm I}$ (\ref{T_h_1}), $\mathcal{T}_{\rm prop}^{\rm II}$ (\ref{T_h_2}), $\mathcal{T}_{\rm sum\_DG}$ (\ref{T_s}), and $\mathcal{T}_{\rm ad\_hoc}$ (\ref{T_c}) using the set $\Omega_1$ with complex Gaussian noise and colored Gaussian noise, respectively, $T_o = 10\,\mu{\rm s}$}\label{ROC_Tx_all}
\end{figure}

The colored Gaussian noise is generated in the same way as in \cite{Huang2013}, where a complex Gaussian noise goes through a three-tap linear filter with coefficients $\{0.3,1,0.3\}$. The parameter $L_n$ is set to $5$ similarly as in \cite{Huang2013}. The detection performance of several detectors ($\mathcal{T}_{\rm prop}^{\rm I}$, $\mathcal{T}_{\rm prop}^{\rm II}$, $\mathcal{T}_{\rm sum\_DG}$, and $\mathcal{T}_{\rm ad\_hoc}$) vs. SNR is compared in Fig.~\ref{Pd_Tx_all}. The detector $\mathcal{T}_{\rm ad\_hoc}$ reviewed in (\ref{T_c}) marginally outperforms the rest in Fig.~\ref{Pd_Tx_all}, as it is dedicated to deal with colored Gaussian noise. The detectors employing the parameter set $\Omega_1$ are better than the ones using $\Omega_2$. Major performance differences exist at the medium SNR range, and the detection performance converge at high SNR.

In order to show more insight about the performance of the detectors, the receiver operating characteristics (ROCs) of different detectors are evaluated in Fig.~\ref{ROC_Tx_all} with ${\rm SNR} = -2~{\rm dB}$ and ${\rm SNR} = 2~{\rm dB}$, respectively. The ROC performance of all detectors under complex Gaussian noise outperforms that under colored Gaussian noise as indicated in Fig.~\ref{ROC_Tx_all}. The higher SNR is, the better ROC performance is. In general, the ROC curves of all the detectors are close to each other. The ROCs of $\mathcal{T}_{\rm sum\_DG}$ and $\mathcal{T}_{\rm ad\_hoc}$ are slightly better than that of the proposed detectors with higher complexity.


\subsection{Multi-cycle multi-lag detection with narrow band interferences under multipath channels}

According to the IEEE 802.15.4a standard \cite{S802.15.4a}, one possible signal occupying the same frequency bands as the UWB signal is from the IEEE 802.16 standard \cite{S802.16} systems (e.g. WiMAX systems). Therefore, the OFDM signals based on the IEEE 802.16 standard are used as a narrow band interference. In this subsection, we evaluate the probability of detecting the UWB signal, which coexists with the OFDM signal under multipath channels.

\begin{table*}[!t]
\centering
\begin{tabular}{c l}
$g(t)$ & the pulse function of length $T_s$ (e.g. the rectangular function)\\
$\zeta$ & the unknown deterministic timing offset    \\
$N_{c}$ & the number of subcarriers, $N_c = 256$ \\
$N_{used}$ & the number of subcarriers used for data, $N_{used} = 200$\\
$\Delta f$ & the carrier separation, $\Delta f = 20\,{\rm MHz}/ N_c = 78.125\,{\rm KHz}$\\
$T_d$ & the length of the data block, where $T_d = 1/\Delta f$ \\
$\rho$ & the ratio of the cyclic prefix block to the data block, $\rho =0.25$ \\
$T_{cp}$ & the length of the cyclic prefix block, $T_{cp} = \rho T_d$\\
$T_{sym}$ & the length of the OFDM symbol, where $T_{sym} = T_d+T_{cp} = (1+ \rho)T_d$ \\
$d_{n,l}$& the $l$th data symbol modulates the $n$th carrier, QPSK modulation, $d_{n,l} \in \{ \pm 1/\sqrt{2} \pm j/\sqrt{2} \}$\\
\end{tabular}
\caption{Parameters for the IEEE 802.16 WiMAX OFDM signal \cite{S802.16}.}\label{OFDM_para}
\end{table*}

According to the IEEE 802.16 standard \cite{S802.16}, the baseband cyclic prefix OFDM signal can be written as
\beqa
\label{OFDM_model}
s(t) &=& \sum_{l= -\infty}^{\infty}\sum_{n=-N_{used}/2, n \neq 0}^{N_{used}/2}d_{n,l}g(t-lT_s - \zeta)\nonumber\\
&&\times e^{j 2\pi n\Delta f(t-lT_s - \zeta)},
\enqa where the parameters in (\ref{OFDM_model}) and their values assigned in the simulations are listed in Table \ref{OFDM_para}. It is well known that the fundamental cyclic frequency $\alpha^s_1$ of the OFDM signal is the symbol rate $1/T_{sym}$, and significant CAF values manifest at the TLs $\pm T_d$ \cite{Oner2007a,Tani2010}. Thus, the OFDM signal has different cyclic features from the UWB signal. Furthermore, the OFDM signal does not have the conjugate cyclic features when it is sampled at the rate $1/(N_cT_d)$. Note that the received signal is sampled at $1$ GHz in order to avoid spectrum aliasing for the UWB signal. Due to this oversampling, the OFDM signal contributes to the conjugate CAF as well. Nevertheless, the estimated conjugate CAFs are still be applied here to calculate the test statistic.

Moreover, the bandwidth of the OFDM signal is $20\,{\rm MHz}$. Its carrier frequency $f_c^\prime$ is randomly generated in the range of $[-240, 240]\,{\rm MHz}$ in each Monte-Carlo run. The OFDM signal goes through a Rayleigh fading channel of $20$ taps with an exponentially decaying power delay profile. The average power of the channel is normalized. The signal-to-interference ratio (SIR) is defined as $\textrm{SIR} = 10{\rm log}_{10}(\sigma_x^2/\sigma_s^2)$, where $\sigma_s^2$ is the power of the OFDM signal. The parameter $L_n$ for $\mathcal{T}_{{\rm ad\_hoc}^\ast}$ is set to $30$ to deal with the OFDM interference as a colored noise.

\subsubsection{Complex Gaussian noise}

\begin{figure}
\centering
\centerline{\includegraphics[width=0.75\linewidth]{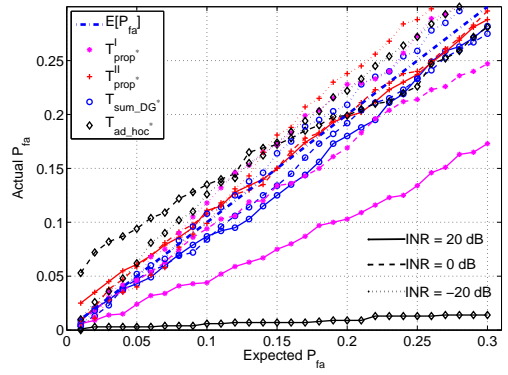}}
\caption{The comparison of the simulated $P_{\rm fa}$ vs. the expected $P_{\rm fa}$ among different detectors $\mathcal{T}_{{\rm prop}^\ast}^{\rm I}$ (\ref{T_h_1}), $\mathcal{T}_{{\rm prop}^\ast}^{\rm II}$ (\ref{T_h_2}), $\mathcal{T}_{{\rm sum\_DG}^\ast}$ (\ref{T_s}), and $\mathcal{T}_{{\rm ad\_hoc}^\ast}$ (\ref{T_c1}) using the parameter set $\Omega_1$, when $T_o = 10\,\mu{\rm s}$, and INR = $\{-20\,{\rm dB}, 0\,{\rm dB}, 20\,{\rm dB}\}$.}\label{Pfa_All}
\end{figure}

%

\begin{figure}
\centering
\centerline{\includegraphics[width=0.75\linewidth]{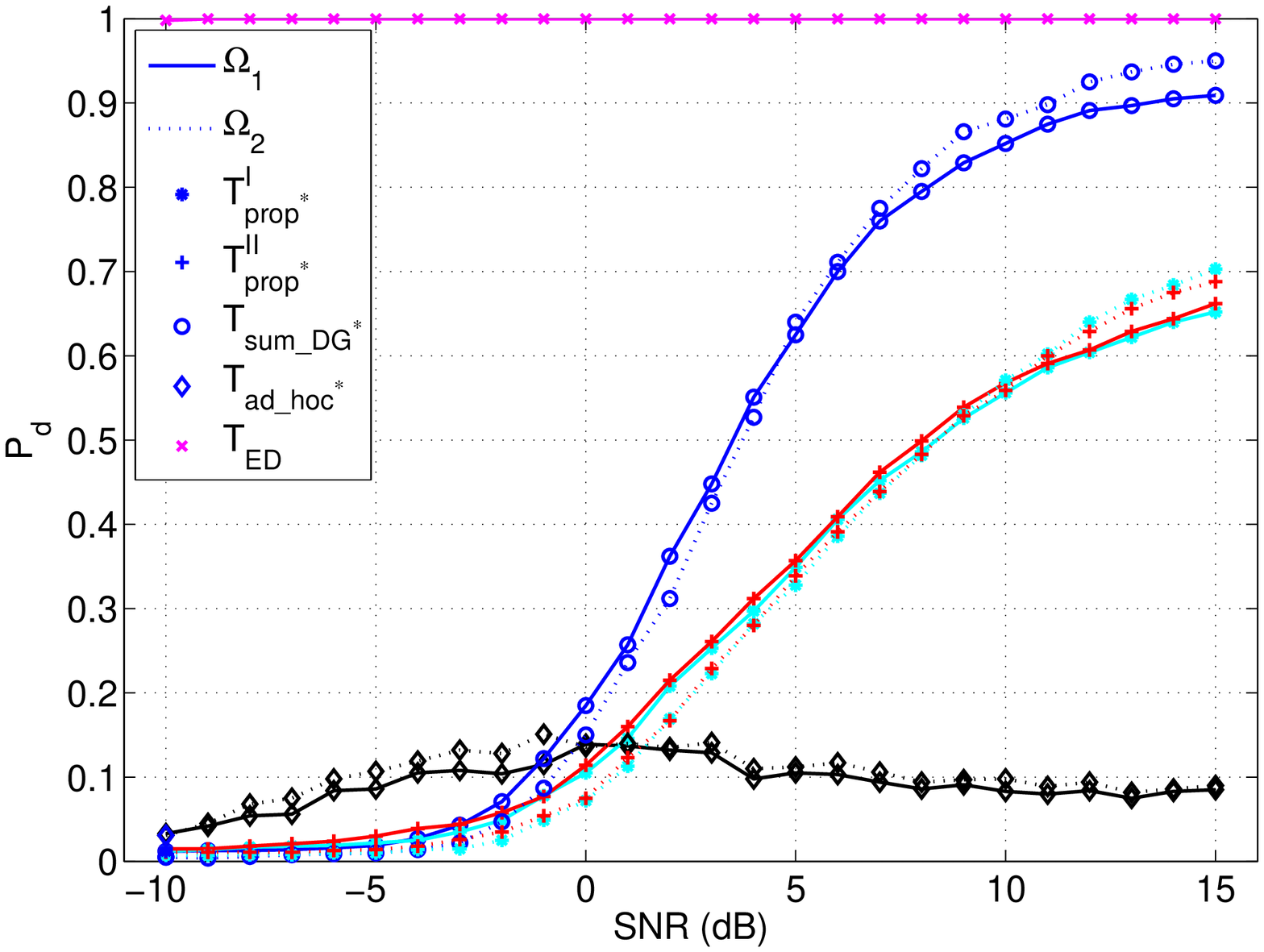}}
\caption{The comparison of $P_{\rm d}$ among the detectors  $\mathcal{T}_{{\rm prop}^\ast}^{\rm I}$ (\ref{T_h_1}), $\mathcal{T}_{{\rm prop}^\ast}^{\rm II}$ (\ref{T_h_2}), $\mathcal{T}_{{\rm sum\_DG}^\ast}$ (\ref{T_s}), $\mathcal{T}_{{\rm ad\_hoc}^\ast}$ (\ref{T_c1}) and $\mathcal{T}_{\rm ED}$ (\ref{T_e}) using different CF and TL sets with the OFDM interference, $T_o = 10\,\mu{\rm s}$, $P_{\rm fa} = 0.01$, ${\rm SIR} = -5\, {\rm dB}$.}\label{Pd_INF}
\end{figure}

\begin{figure}
\centering
\centerline{\includegraphics[width=0.75\linewidth]{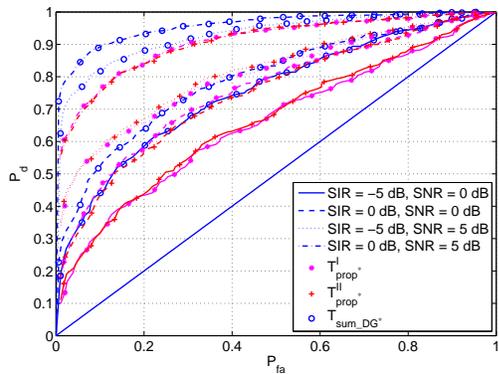}}
\caption{The comparison of ROC among the detectors  $\mathcal{T}_{{\rm prop}^\ast}^{\rm I}$ (\ref{T_h_1}), $\mathcal{T}_{{\rm prop}^\ast}^{\rm II}$ (\ref{T_h_2}), $\mathcal{T}_{{\rm sum\_DG}^\ast}$ (\ref{T_s}), $\mathcal{T}_{{\rm ad\_hoc}^\ast}$ (\ref{T_c1}) and $\mathcal{T}_{\rm ED}$ (\ref{T_e}) using the parameter set $\Omega_1$ with the OFDM interference, $T_o = 10\,\mu{\rm s}$.}\label{ROC_INF}
\end{figure}

As the OFDM interference is just in the band of interest, there may be circumstances that only the OFDM interference and the complex Gaussian noise exist. Under such circumstance, the actual $P_{\rm fa}$ may be different from the expectation. Let us define the interference-to-noise ratio (INR) as $\textrm{INR} = 10{\rm log}_{10}(\sigma_s^2/\sigma_n^2)$. Consequentially, the difference between the actual $P_{\rm fa}$'s (the simulated ones) and the expected $P_{\rm fa}$'s of various detectors using the conjugate CAF estimates is explored in Fig.~\ref{Pfa_All}. The parameter set $\Omega_1$ is employed. When INR = $-20\,{\rm dB}$, the noise dominates. The actual $P_{\rm fa}$ values of the detectors (the dotted lines) generally follow the expected $P_{\rm fa}$ (the dash-dotted line) values. However, they deviate to some extent from the expectation, when the expected $P_{\rm fa}$ increases. When INR increases to $0\,{\rm dB}$, the actual $P_{\rm fa}$ of the detector $\mathcal{T}_{{\rm ad\_hoc}^\ast}$ is larger than the expectation obviously in the range of small expected $P_{\rm fa}$'s. It converges to the expectation after the expected $P_{\rm fa} = 0.2$. Moreover, the actual $P_{\rm fa}$ of the detector $\mathcal{T}_{{\rm prop}^\ast}^{\rm I}$ is slightly smaller the expectation, which is harmless. The $P_{\rm fa}$'s of other detectors are still around the predesigned values. When INR = $20\,{\rm dB}$, the interference dominates. The actual $P_{\rm fa}$'s of $\mathcal{T}_{{\rm prop}^\ast}^{\rm II}$ and $\mathcal{T}_{{\rm sum\_DG}^\ast}$ closely follow the benchmark. On the other hand, the actual $P_{\rm fa}$'s of $\mathcal{T}_{{\rm prop}^\ast}^{\rm I}$ and $\mathcal{T}_{{\rm ad\_hoc}^\ast}$ are below the benchmark, thus the thresholds are overestimated. The $P_{\rm d}$'s would decrease as well. Especially, the $P_{\rm fa}$ curve of $\mathcal{T}_{\rm ad\_hoc^\ast}$ is almost flat and far below the benchmark. The degradation of its detection performance would be obvious.

Now let us evaluate the detection performance under the OFDM interference and complex Gaussian noise. The detection performance of various detectors are indicated in Fig.~\ref{Pd_INF}, where $\textrm{SIR} = -5\,{\rm dB}$. The OFDM interference dramatically changes the performance of the detectors. The detector $\mathcal{T}_{{\rm ad\_hoc}^\ast}$ is the worst. It fails to detect the signal of interest due to its oversimplified covariance estimation. This result is consistent with the prediction based on Fig~\ref{Pfa_All}. Regardless of the correlation between different CFs, the detector $\mathcal{T}_{{\rm sum\_DG}^\ast}$ exploits the correlation of $\hat{R}_{xx^\ast}(\alpha_i)$ using the same CF but different TLs, and it performs best. The proposed detectors ($\mathcal{T}_{{\rm porp}^\ast}^{\rm I}$ and $\mathcal{T}_{{\rm prop}^\ast}^{\rm II}$) neglect the correlation between $\hat{R}_{xx^\ast}(\alpha_i,\tau_{i,l})$ of different CF-TL pairs, and thus suffer from the performance degradation. It trades the detection performance with the low complexity. Moreover, the energy detector $\mathcal{T}_{\rm ED}$ fails to distinguish the interference and the signal of interest, and thus its $P_{\rm d}$ is always $1$ by mistakenly regarding the interference as the signal.

In Fig.~\ref{ROC_INF}, the ROC is investigated under different combinations of SIR and SNR using the parameter set $\Omega_1$. The higher SIR and SNR are, the higher probability of detection is. The detection performance $P_{\rm d}$ benefits more from the higher SNR than the higher SIR. The detector $\mathcal{T}_{{\rm sum\_DG}^\ast}$ is obviously superior to the proposed detectors in good agreement with Fig.~\ref{Pd_INF}.

\subsubsection{Colored Gaussian noise}

\begin{figure}
\centering
\centerline{\includegraphics[width=0.75\linewidth]{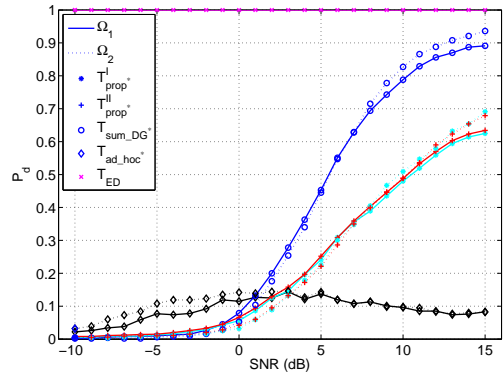}}
\caption{The comparison of $P_{\rm d}$ among the detectors $\mathcal{T}_{{\rm prop}^\ast}^{\rm I}$ (\ref{T_h_1}), $\mathcal{T}_{{\rm prop}^\ast}^{\rm II}$ (\ref{T_h_2}), $\mathcal{T}_{{\rm sum\_DG}^\ast}$ (\ref{T_s}), $\mathcal{T}_{{\rm ad\_hoc}^\ast}$ (\ref{T_c1}) and $\mathcal{T}_{\rm ED}$ (\ref{T_e}) using different CF and TL sets with OFDM interference under colored Gaussian noise, $T_o = 10\,\mu{\rm s}$, $P_{\rm fa} = 0.01$, ${\rm SIR} = -5\,{\rm dB}$.}\label{Pd_INF_CN}
\end{figure}

In the case of the OFDM interference and the colored Gaussian noise, the performance of detectors further degrades as shown in Fig.~\ref{Pd_INF_CN}, where the colored Gaussian noise is generated in the same way as in Section~\ref{sec:CN}. Different detectors behave in the similar tendency as the ones in Fig.~\ref{Pd_INF}. Note that the detector $\mathcal{T}_{{\rm ad\_hoc}^\ast}$ designed for the colored noise cannot detect due to the OFDM interference. Since the ROC performance of the proposed detectors and the detector $\mathcal{T}_{{\rm sum\_DG}^\ast}$ under colored Gaussian noise maintains the same properties as the one under complex Gaussian noise, it is not shown in this paper to save the space.

\section{Conclusions}\label{sec:concl}

In this paper, we propose multi-cycle multi-lag cyclic feature detectors for UWB receivers in heterogeneous environments. The unique cyclic features of UWB signals based on the IEEE 802.15.4a standard are analyzed. Due to the ultra wide bandwidth, the UWB signal manifests itself at several cyclic frequencies. Furthermore, the multipath channel effects increase the range of its cyclic features w.r.t. the time lag. The constant false alarm rate detectors based on cyclic features are proposed accordingly. Their computational complexities are significantly less than the ones of the conventional Dandawate-Giannakis detector and its variations. Extensive simulation results indicate that the proposed detectors introduce tradeoffs between the detection performance and the computational complexity in various scenarios, such as multipath channels, colored Gaussian noise and OFDM interferences.

\appendices

\section{Derivation of $R_{xx}(t,\tau)$}\label{app:AF}
Making use of the Fourier transform pair $p(t) = \int_{-\infty}^{\infty} P(f)e^{j2\pi f t}df$, where $P(f)$ is the Fourier transform of $p(t)$, the signal $x(t)$ (\ref{UWB_model}) can be written as
\beqa\label{UWB_model2}
x(t)
\!=\! \!\!\sum_{k=-\infty}^{+\infty}\!\!a_k\!\! \sum_{n=0}^{N_{cpb}-1}\!\!c_{n+kN_{cpb}}\int \left\{ P(f) e^{j2\pi f(t-\epsilon-nT_c)}\right.\nonumber\\
\left. \times e^{-j2\pi fkT_{dsym}} e^{-j2\pi fb_kT_{BPM}} e^{-j2\pi fh^{(k)}T_{burst}}  \right\} df.
\enqa  Since $c_{k}$ and $a_k$ take values from $\{\pm 1\}$ with equal probability, we obtain that $E[a_ka_{k-l}^\ast]=\delta(l)$ and $E[c_kc_{k-l}^\ast]=\delta(l)$, where $\delta(l)$ is the delta function. Furthermore, let us define $\bar{\beta}(f) = E[ e^{-j2\pi f b_k T_{BPM}}]$ and $\bar{\eta}(f) = E[e^{-j2\pi f h^{(k)} T_{burst}}]$. Using these definitions, and plugging (\ref{UWB_model2}) into (\ref{cov_nc}), we arrive at
\beqa
&&\Upsilon_{xx}(t,\tau)\label{Rx1}\\
&\!=\!& \sum_{k=-\infty}^{+\infty}\!\!\sum_{n=0}^{N_{cpb}-1}\int \int \left\{ P(y)P^\ast(z)  \bar{\beta}(y-z)\bar{\eta}(y-z)\right. \nonumber\\
&& \left.\times e^{j2\pi((y-z)(t-\epsilon) + \tau (y+z)/2 - T_{dsym}(y-z)k - T_c(y- z)n)}\!\!\right\} dy dz.\nonumber
\enqa According to the Poisson sum formula $\displaystyle\sum_{k=-\infty}^{+\infty}e^{-j2\pi k fT} = \frac{1}{T}\sum_{q=-\infty}^{+\infty}\delta(f-\frac{q}{T})$,  and denoting $\alpha_q^x = q/T_{dsym}, q \in \mathcal{Z}$, the equation (\ref{Rx1}) can be further rewritten w.r.t. $k$ summations as (\ref{ACF_UWB}).


\section{Derivation of $R_{xx}(\alpha_q^x,\tau)$}\label{app:CAF}
Recalling that $T_{burst} = N_{cpb}T_c$ and $T_{dsym} = N_{burst}T_{burst}$ in Table~\ref{UWB_para}, the CAF $R_{xx}(\alpha_q^x,\tau)$ as the Fourier coefficient of $\Upsilon_{xx}(t,\tau)$ is given by
\beqa\label{Ra}
R_{xx}(\alpha_q^x,\tau)
&\!\!\!\!=\!\!\!\!&\alpha_1^x e^{-j2\pi \alpha_q^x\epsilon}\bar{\beta}(\alpha_q^x)\bar{\eta}(\alpha_q^x) \phi_p(\alpha_q^x,\tau)\nonumber\\
&&\times w\left(q/(N_{burst}N_{cpb}),N_{cpb}\right).
\enqa Note that $b_k$ and $h^{(k)}$ select values from $\{0,1\}$ and $\{0,1,\dots, N_{hop}-1\}$ with equal probability, respectively, and they are independent with each other. Making use of $T_{BPM}=T_{dsym}/2$ in Table~\ref{UWB_para}, we arrive at
\beqa
\label{beta}\bar{\beta}(\alpha_q^x)&\!=\!&\frac{1+e^{-j2\pi T_{BPM}\alpha_q^x}}{2}  \nonumber\\
&\!=\!&\frac{1+(-1)^q}{2} = \left\{\begin{array}{cc}
1 & q = 0, \pm 2,\pm 4, \dots \\
0 & q = \pm 1,\pm 3, \dots
\end{array} \right., \\
\label{eta}\bar{\eta}(\alpha_q^x)&\!=\!&\frac{1}{N_{hop}}\sum_{n=0}^{N_{hop}-1}e^{-j2\pi  T_{burst}\alpha_q^x n}\nonumber\\
&\!=\!&\frac{1}{N_{hop}}w\left(q/N_{burst},N_{hop}\right).
\enqa As $\bar{\beta}(\alpha_q^x)$ is nonzero only when $q$ is an even number, the fundamental CF is doubled as $2\alpha_1^x$. For the sake of brevity, we abuse $q \in \mathcal{Z}$. Plugging (\ref{beta}) and (\ref{eta}) into (\ref{Ra}), the CAF $R_{xx}(2\alpha_q^x,\tau)$ of $x(t)$ can be written as (\ref{CAF_UWB}).

%

\bibliographystyle{IEEEbib}
{\footnotesize\bibliography{IEEEabrv,ICUWB13}}

\end{document}

%% file: ajmacros.tex
\makeatletter
\def\argmin{\mathop{\operator@font argmin}}
\def\argmax{\mathop{\operator@font argmax}}
\makeatother

\newcommand{\bea}{\begin{array}}
\newcommand{\ena}{\end{array}}
\newcommand{\beq}{\begin{equation}}
\newcommand{\enq}{\end{equation}}

\newcommand{\beqa}{\begin{eqnarray}}
\newcommand{\enqa}{\end{eqnarray}}

\newcommand{\beqan}{\begin{eqnarray*}}
\newcommand{\enqan}{\end{eqnarray*}}

\newcommand{\AL}{\begin{enumerate}}
\newcommand{\ALE}{\end{enumerate}}

\def\addots{\mathinner{
    \mkern1mu\raise0pt\vbox{\kern7pt\hbox{.}}
    \mkern2mu\raise4pt\hbox{.}
    \mkern2mu\raise7pt\hbox{.}
    \mkern1mu}}

\def\sddots{\mathinner{
    \mkern.8mu\raise7pt\hbox{.}
    \mkern.8mu\raise4pt\hbox{.}
    \mkern.8mu\raise0pt\vbox{\kern7pt\hbox{.}}
    \mkern1mu}}

\def\saddots{\mathinner{
    \mkern.2mu\raise0pt\vbox{\kern7pt\hbox{.}}
    \mkern.2mu\raise4pt\hbox{.}
    \mkern.2mu\raise7pt\hbox{.}
    \mkern1mu}}

\newcommand{\br}{{\ensuremath{\mathbf{r}}}}

\newcommand{\bP}{{\ensuremath{\mathbf{P}}}}
\newcommand{\bQ}{{\ensuremath{\mathbf{Q}}}}

\makeatletter
\def\setboxz@h{\setbox\z@\hbox}
\def\wdz@{\wd\z@}
\def\boxz@{\box\z@}
\def\underset#1#2{\binrel@{#2}%
  \binrel@@{\mathop{\kern\z@#2}\limits_{#1}}}
\def\binrel@#1{\begingroup
  \setboxz@h{\thinmuskip0mu
    \medmuskip\m@ne mu\thickmuskip\@ne mu
    \setbox\tw@\hbox{$#1\m@th$}\kern-\wd\tw@
    ${}#1{}\m@th$}%
  \edef\@tempa{\endgroup\let\noexpand\binrel@@
    \ifdim\wdz@<\z@ \mathbin
    \else\ifdim\wdz@>\z@ \mathrel
    \else \relax\fi\fi}%
  \@tempa
}
\let\binrel@@\relax%
\makeatother

\def\circone{
    {\ooalign{
   \smash{\hskip3.0pt\raise0pt\hbox{\small $1$}}\vphantom{}\crcr
   \hbox{$\bigcirc$}
    }}
    }
\def\circtwo{
    {\ooalign{
   \smash{\hskip3.0pt\raise0pt\hbox{\small $2$}}\vphantom{}\crcr
   \hbox{$\bigcirc$}
    }}
    }
\def\circthree{
    {\ooalign{
   \smash{\hskip3.0pt\raise0pt\hbox{\small $3$}}\vphantom{}\crcr
   \hbox{$\bigcirc$}
    }}
    }

\def\sqplus{\mathbin{
    {\ooalign{\hfil\raise.3ex\hbox{\scriptsize
    +}\hfil\crcr\mathhexbox274\crcr\mathhexbox275}}
    }}
\def\sqminus{\mathbin{
    {\ooalign{\hfil\raise.3ex\hbox{\scriptsize
    --}\hfil\crcr\mathhexbox274\crcr\mathhexbox275}}
    }}

\def\IC{{
   \mathord{
      \hbox to 0em{
     \hskip-4pt
         \ooalign{
       \smash{\hskip1.9pt\raise2.6pt\hbox{$\scriptscriptstyle |$}}\crcr
       \smash{\hbox{\rm\sf C}} }
     \hidewidth}
      \phantom{\hbox{\rm\sf C}}
} }}
\def\IN{
    {\ooalign{
   \smash{\hskip2.2pt\raise1.5pt\hbox{$\scriptscriptstyle |$}}\vphantom{}\crcr
   \hbox{\sf N}
    }}
    }
\def\IZ{
    {\ooalign{
   \smash{\hskip1.9pt\raise0pt\hbox{$\sf Z$}}\vphantom{}\crcr
   \hbox{\sf Z}
    }}
    }
\def\IR{
    {\ooalign{
   \smash{\hskip2.2pt\raise1.5pt\hbox{$\scriptscriptstyle |$}}\vphantom{}\crcr
   \smash{\hskip2.2pt\raise3.3pt\hbox{$\scriptscriptstyle |$}}\vphantom{}\crcr
   \hbox{\sf R}
    }}
    }

\iffalse
    \DeclareMathAlphabet{\mathcmb}{OT1}{cmr}{bx}{n}

\else
    \usepackage{amsbsy}

\fi

\newcommand{\SI}{\begin{indlist}}
\newcommand{\EI}{\end{indlist}}

\newcommand{\DL}{\begin{dashlist}}
\newcommand{\DLE}{\end{dashlist}}